\documentclass[twocolumn]{aastex62}
\usepackage[utf8]{inputenc}
\usepackage[colorinlistoftodos]{todonotes}
\usepackage{float}
\usepackage{amsmath}
\hypersetup{breaklinks,colorlinks,urlcolor=blue,citecolor=blue,linkcolor=blue}

\renewcommand{\arraystretch}{0.5}

\begin{document}
\title{{\bf Neutron Star Mergers as the Main Source of \textsl{r}-process: Natal Kicks And Inside-Out Evolution to The Rescue}}
\author[0000-0002-6389-2697]{Projjwal Banerjee}
\email{projjwal.banerjee@gmail.com}
\affil{Discipline of Physics, Indian Institute of Technology Palakkad, Kerala, India
678557;~\href{mailto:projjwal.banerjee@gmail.com}{\rm{projjwal.banerjee@gmail.com}}}

\author[0000-0003-4960-8706]{Meng-Ru Wu}
\affil{Institute of Physics, Academia Sinica, 
  Taipei, 11529, Taiwan;~\href{mailto:mwu@gate.sinica.edu.tw}{\rm{mwu@gate.sinica.edu.tw}}}
\affil{Institute of Astronomy and Astrophysics, Academia Sinica, 
  Taipei, 10617, Taiwan} 
\affil{Physics Division, National Center for Theoretical Sciences, Hsinchu, 30013, Taiwan}
  
\author[0000-0002-8129-5415]{Zhen Yuan}
\affil{Key Laboratory for Research in Galaxies and Cosmology, Shanghai Astronomical Observatory, Chinese Academy of Sciences, 80 Nandan Road, Shanghai 200030, China;~\href{sala.yuan@gmail.com}{\rm{ sala.yuan@gmail.com}}}
\affil{Universit\'e de Strasbourg, CNRS, Observatoire astronomique de Strasbourg, UMR 7550, F-67000 Strasbourg, France}

\date{\today}

\begin{abstract}
Binary neutron star mergers (BNSMs) are currently the most promising source of \textsl{r}-process thanks to the detection of GW170817. The estimated occurring frequency and the amount of mass ejected  per merger indicate that BNSMs by themselves can account for all the \textsl{r}-process enrichment in the Galaxy. 
However, the decreasing trend of [Eu/Fe] versus [Fe/H] of disk stars for [Fe/H]$\gtrsim -1$ in the solar neighborhood is inconsistent with the flat trend expected from BNSMs with a standard delay time distribution (DTD) $\propto t^{-1}$.
This has led to the suggestion that either additional sources or modification to the DTD of BNSMs is required to match the observations. 
We investigate the effects of natal kicks received during the birth of neutron star binaries on the chemical evolution of \textsl{r}-process element Eu in the Milky Way by combining the results from the galactic dynamics code \textsc{galpy} with a one-zone Galactic chemical evolution model \textsc{omega}.
We show that when key inputs from simulations of the inside-out disk evolution are combined with natal kicks, BNSMs can naturally reproduce the observed decreasing trend of [Eu/Fe] with [Fe/H] in the solar neighborhood without the need for modification to the DTD or additional \textsl{r}-process sources.

\end{abstract}

\section{Introduction}

Following the seminal discovery of GW170817 \citep{abbott2017a,abbott2017b}, binary neutron star mergers (BNSMs) have become the first, and currently the only, confirmed site for the synthesis of heavy elements by the rapid neutron capture process (\textsl{r}-process; see \citet{Cowan2019,Metzger2019} for recent reviews). The current estimated BNSM rate of $110$--$3840~{\rm Gpc^{-3}~yr^{-1}}$ \citep{Ligo2019} along with the amount of \textsl{r}-process material ejected per merger of  $\approx 0.03$--$0.06~M_{\odot}$ (e.g.~\citealt{Drout:2017ijr,Cowperthwaite+17,Kasen+17,Tanaka+2017,Villar+17,Kawaguchi+18,Wanajo2018,Wu+19}) is sufficient to explain all of the \textsl{r}-process enrichment in the Galaxy \citep{cote2018,hotokezaka+2018,Cowan2019}. However, if BNSMs are assumed to be the only source of \textsl{r}-process, then the Galactic evolution of elements such as Eu, which are primarily produced by \textsl{r}-process, is very different from elements that are produced by other sources associated with massive stars. This is due to the inherent delay between the birth of a neutron star binary (NSB) and the eventual merger with a typical delay time distribution (DTD) $\propto t^{-1}$, compared to massive stars that have negligible delay. 
In particular, with a DTD $\propto t^{-1}$ that is expected from population synthesis calculations \citep{dominik+2012,chruslinska+2018,Cote2019}, [Eu/Fe] remains almost constant for stars with [Fe/H]$\,\gtrsim -1$ in galactic chemical evolution (GCE) calculations due to the fact that both BNSMs and Type Ia supernovae (SNe Ia) have the same DTD \citep{vandevoort+2015,Komiya2016}\footnote{We note that the decreasing trend of [Eu/Fe] at high metallicity can be reproduced if a fixed delay time for BNSM is adopted. See, e.g.,~\citet{argast2004,Matteucci2014,Wehmeyer2015}.}.
In sharp contrast, the observed values of [Eu/Fe] in disk stars at the solar neighborhood show a clearly decreasing trend  with increasing [Fe/H] \citep{battistini2016}. This was pointed out to be an issue by \citet{cote2017a} and \citet{hotokezaka+2018} and has been studied in detail recently by \citet{Cote2019} \citep[see also][]{ Schonrich2019}. 
Possible solutions to the problem include a steeper DTD ($\propto t^{-1.5}$) or a burst of merger at early times followed by a typical DTD $\propto t^{-1}$ (see also \citealt{hotokezaka+2018}). 
These, however, have been pointed out to be inconsistent with observations of short gamma ray bursts and SNe Ia in early-type galaxies (see \citealt{Cote2019} for details). 
Alternatively, additional source(s) of \textsl{r}-process can explain the observed trend provided that their frequency decreases with metallicity. This source could be an \textsl{r}-process site associated with the death of massive stars such as magnetorotational SNe \citep{Winteler+12,Mosta+17}, accretion disk outflow from collapsars \citep{Siegel+18}, or supernova explosions associated with hadron-quark phase transition \citep{FischerWu2020}. 
Higher frequency of mergers resulting from neutron star--black hole binaries in the early Galaxy (lower metallicites; \citet{Mennekens2014}) and \textsl{r}-process occurring in accretion disk outflows during the common envelope phase of NS--massive star system \citep{Aldana2019} are some of the other possibilities.
Such sources may also be needed to explain the presence of \textsl{r}-process elements in the very early Galaxy and the large scatter in their abundances as observed in very metal-poor stars with [Fe/H]$\lesssim-2.5$~\citep{argast2004,Wehmeyer2019} but this is still under debate~\citep{Tsujimoto2014,Hirai2015,Ishimaru+2015,Shen2015,vandevoort+2015,Safarzadeh+2019}. 

An interesting feature that distinguishes BNSMs from other rare \textsl{r}-process sites is that they receive large natal kicks during the formation of binaries (\citet{Fong&Berger13,behroozi2014}; see, however, \citet{BeniaminiPiran2016,Tauris2017}).  
Consequently, the final location of mergers can be very different from their birth location where a certain fraction of NSBs are effectively lost as they do not contribute to the Galactic enrichment of heavy elements due to the fact that they merge far away from the star forming regions ~\citep{safar2017,Safarzdeh+2017}. In this Letter, we show that  natal kicks have a large impact even for NSBs that do contribute to the Galactic enrichment of heavy elements. Specifically, we model two new effects due to natal kicks on GCE; the effect of kick-induced migration on the effective BNSM frequency as well as the impact on the effective DTD. We show that when these effects are combined with the inside-out formation of the Milky Way (MW; see e.g., \citet{minchev2013,Schonrich17,frankel19}), the decreasing trend of [Eu/Fe] versus [Fe/H] for stars with [Fe/H]$\gtrsim -1$ matching the observation can be naturally obtained with BNSMs as the only \textsl{r}-process source with a standard DTD $\propto t^{-1}$.

\section{Effects of Natal Kicks}\label{sec:kick}
It is known, that due to natal kicks, only a fraction $f_{\rm ret}\leq 1$ of the total NSBs born at a certain time interval contribute to the Galactic enrichment of heavy elements, whereas the rest merge far from the star forming regions \citep{safar2017}. In addition, however, there are two other important effects caused by natal kicks that have not been taken into account previously but turn out to be crucial in modeling the GCE of \textsl{r}-process elements in the solar neighborhood. The first one is the kick-induced migration of NSBs within the Galactic disk. Considering a simple model of the MW disk as consisting of independent concentric rings, one can study the GCE in the vicinity of a particular radius that is described by a ring centered at the given radius. For any ring, the GCE of non-\textsl{r}-process elements that receive negligible contribution from BNSMs depends mostly on the local star formation rate (SFR). In contrast, for \textsl{r}-process elements, their GCE in a given ring depends not only on the BNSMs that are both born and merge inside the ring, but also on BNSMs that are born in other rings that migrate and eventually merge within the considered ring. In this regard, we define a useful quantity relevant for GCE calculations, $\eta(R,t)$, as
\begin{equation}
    \eta(R,t)=\frac{N_{\rm merge}(R,t)}{N_{\rm born}(R,t)}
    \label{eq:eta}
\end{equation}
where $N_{\rm born}(R,t)$ is the number of NSBs born inside the ring centered at radius $R$ between time $t$ and $t+\Delta t$, and  
$N_{\rm merge}(R,t)$ is the actual number of NSBs born in the entire disk during the same time interval but eventually merge within the ring centered at $R$. If migration due to natal kicks is neglected, only the NSB that is born inside a ring can merge within that ring such that $\eta$ is 1 for all rings. 
It is important to note that although $\eta(R,t)$ effectively alters the birth rate of BNSMs at a time $t$, the mergers occur later according to the DTD.

The second important effect of natal kicks is that it impacts the effective DTD of BNSMs for a given ring. 
This is simply due to the fact that NSBs with shorter merger times tend to be retained and coalesce within the star forming region of the Galaxy, whereas those with longer merger times have a higher chance of escaping. This leads to lower values of average merger times and thus results in an effective value of $\beta^{\rm eff}(R,t)\leq \beta$ for an actual DTD $\propto t^{\beta}$. 

\begin{figure*}
\centerline{\hspace*{0.cm}\includegraphics[width=89mm]{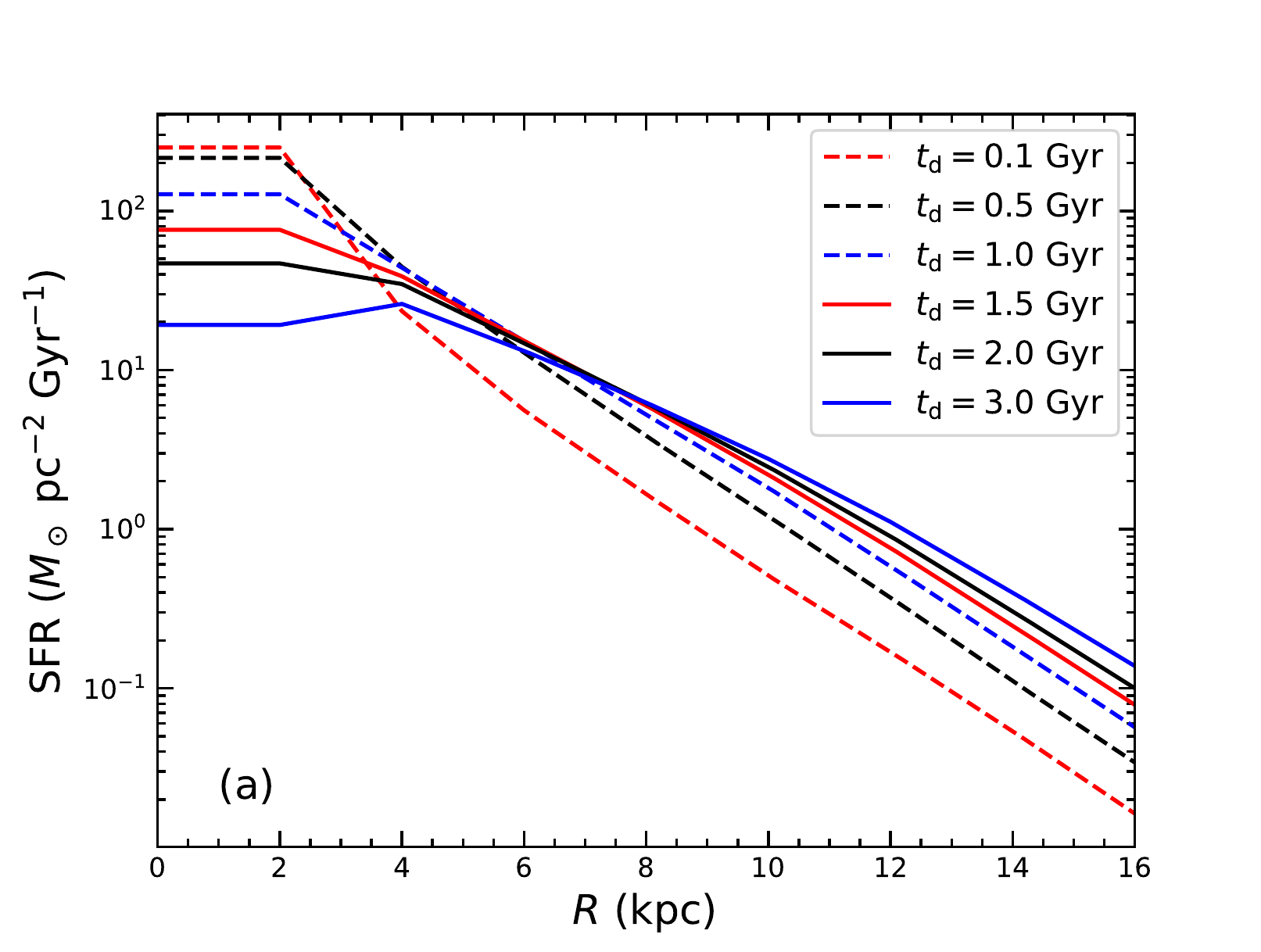}\vspace*{.00cm}\hspace*{-.25cm}\includegraphics[width=79mm]{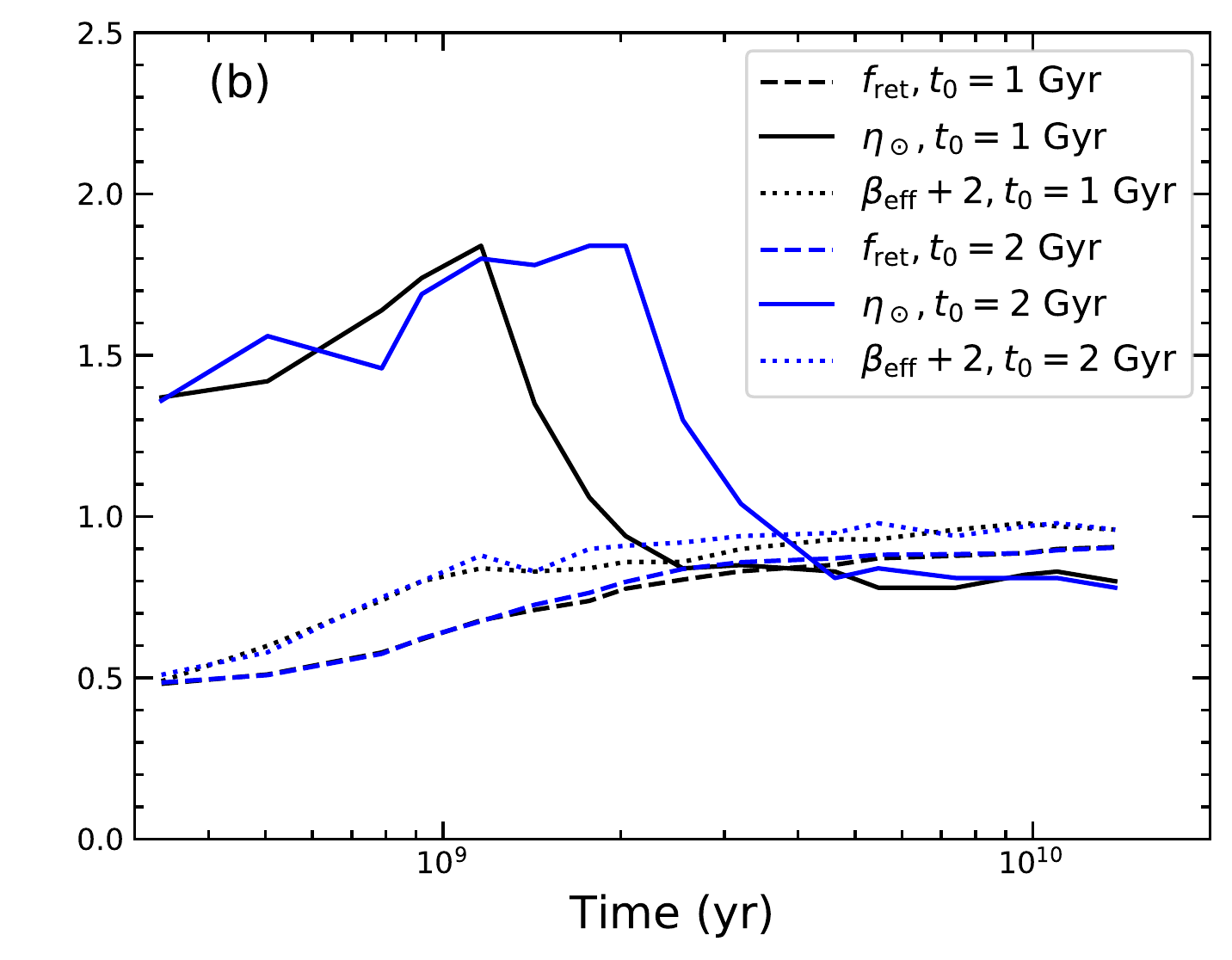}}
\centerline{\hspace*{.40cm}\includegraphics[width=78mm]{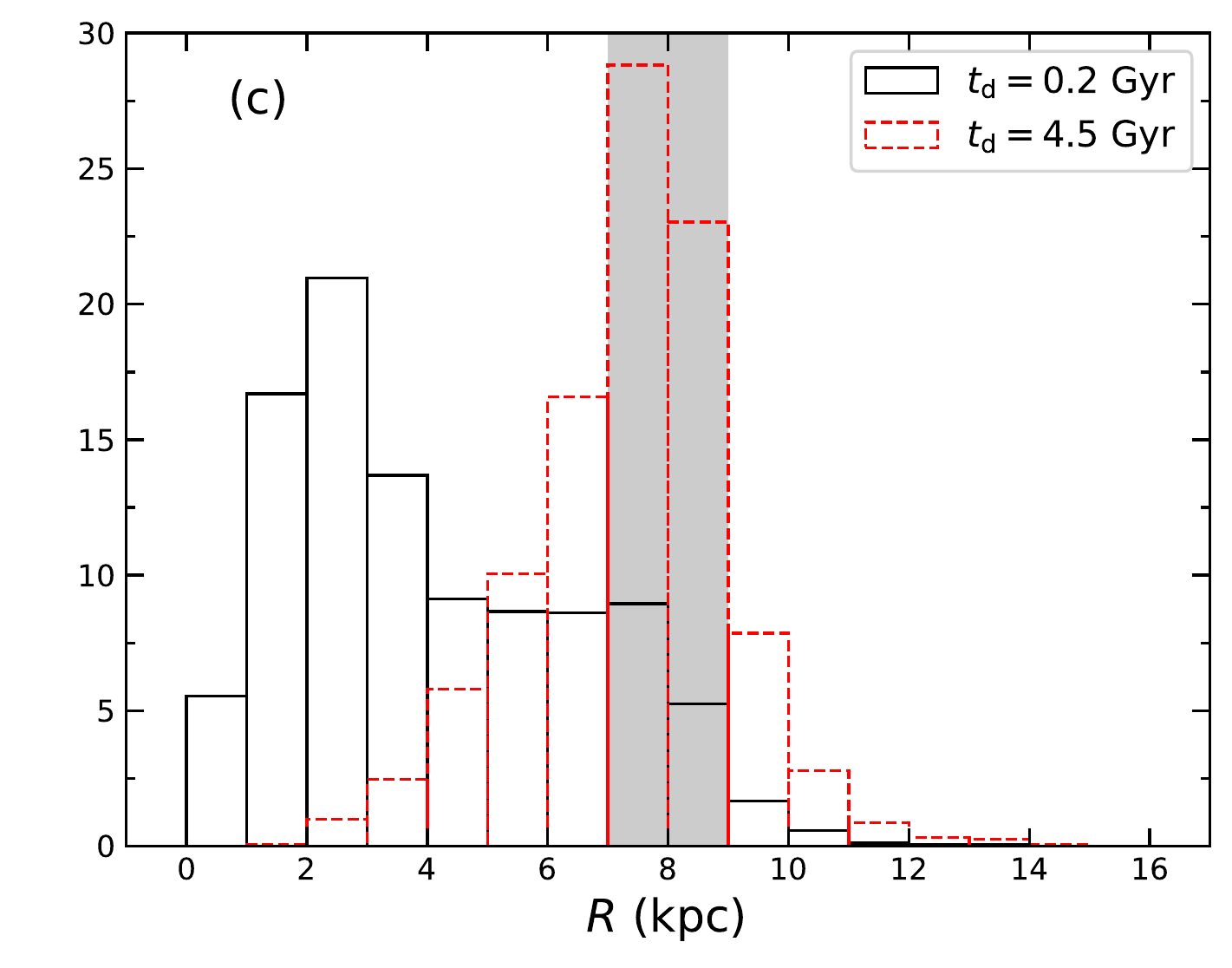}\hspace*{.50cm}\includegraphics[width=79mm]{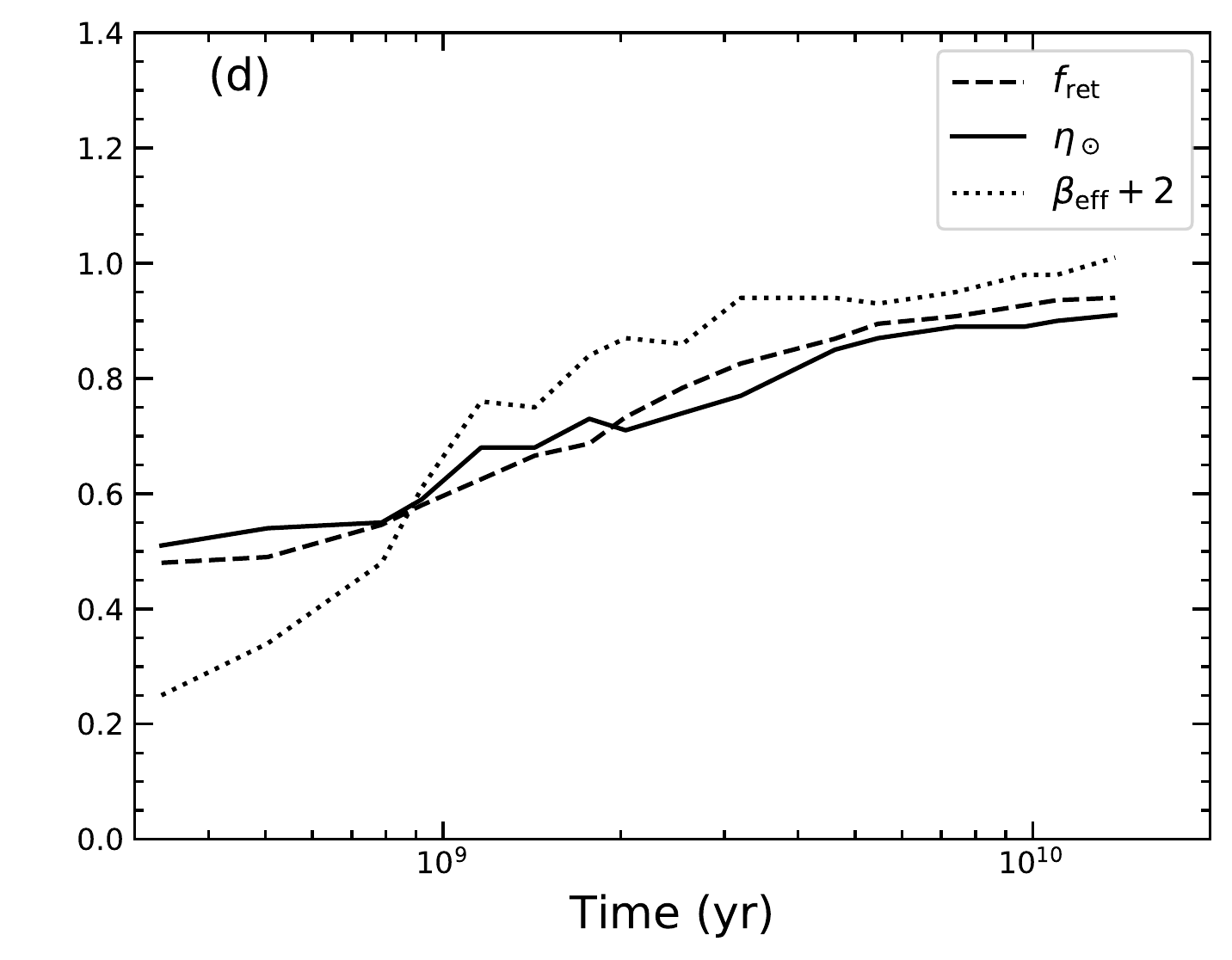}}
\caption{(a) SFR as a function of $R$ from \citet{minchev2013} at various stages of disk evolution. 
(b) Evolution of $f_{\rm ret}, \eta_{\odot}$ and $\beta^{\rm eff}_\odot$ for $t_0=1$ and $2$ Gyr. 
(c) Distribution of the birth radius of NSBs that merge within the solar neighborhood (shown in gray) at two different stages of the disk for $t_0=1$ Gyr. The total number of NSBs at each time is normalized to 100. (d) Same as (b), but using a fixed radial scale length of 3 kpc (no inside-out formation).}
\label{fig:minchev}
\end{figure*}

As mentioned above, the values of $\eta(R,t)$ and $\beta^{\rm eff}(R,t)$  for a certain vicinity (ring) are influenced by the birth rate of NSBs born both inside and outside the ring. Consequently, in order to calculate their values, it is important to know 
both the spatial and temporal evolution of the SFR of the entire disk. 
We use the SFR predicted by a detailed chemodynamical simulation based on the inside-out formation of the MW disk from \citet{minchev2013}.
At any given time $t$, we generate the radial coordinate of the birth locations $R_{\rm b}$ of the NSBs according to a distribution  $\propto R\,\Sigma(R,t)$, where $\Sigma(R,t)$ is the surface SFR density  adapted from \citet{minchev2013}, shown in Fig.~\ref{fig:minchev}(a).
Because the starting time, $t_0$, from the simulation in \citet{minchev2013} is when the bulge is formed, we adopt two different values of $t_0=1$ and $2$~Gyr.
For $t<t_0$, we keep the radial dependence of the SFR the same as that at $t=t_0$. 
The maximum value of $R_{\rm b}$ for the birth location of NSBs is limited to $16$ kpc corresponding to the maximum value for which SFR is provided in \citet{minchev2013}. For simplicity, we assume that all NSBs are born at a vertical height of $z=0$. 

For each NSB born at a given $R$ and $t$, we assign a kick velocity $\vec v_{\rm kick}$ whose magnitude is randomly generated from an exponential distribution $\propto \exp(-v/v_0)$, with $v_0=90~{\rm km\,s^{-1}}$ similar to \citet{behroozi2014}, which is consistent with the kick velocities inferred from the observed offsets of short gamma-ray bursts by \citet{Fong&Berger13}. 
As the latter study inferred a nonzero value for the lower limit of $v_{\rm kick}$, we assume a minimum value of $10~{\rm km\,s^{-1}}$. The direction of $\vec v_{\rm kick}$ is generated from a uniform and isotropic distribution. 
The randomly sampled $\vec v_{\rm kick}$ is then added to the velocity of the NSB (just before the birth of the second neutron star), which is assumed to be the circular velocity corresponding to its birth radius. 

In order to follow the motions of NSBs under the influence of the Galactic potential until they merge, we use \textsc{galpy} \citep{galpy} to trace their orbits. Each NSB is evolved in time until $t_{\rm merge}$, sampled from DTD $\propto t^{-1}$ with minimum and maximum values of 10 Myr and 10 Gyr, respectively. Because \textsc{galpy} only allows for a static potential, we restrict values of $t_{\rm merge}\leq2$ Gyr to minimize the effect of disk growth on the motion of NSB. We use \texttt{MWPotential2014} in \textsc{galpy} as the model for the MW potential with the default value of the circular velocity $v_c(R_\odot$ = 8 kpc)=220~${\rm km\,s^{-1}}$ at the present time. In order to account for the time evolution of the Galactic potential, we assume that the MW potential is proportional to the virial mass $M_{\rm vir}$ of the dark matter halo. Thus, for any given $t$, we simply scale the potential by changing the value of $v_c(R_\odot,t)$ given by 
\begin{equation}
        v_c(R_\odot,t)=v_c(R_\odot,t_{\rm gal})\left( \frac{M_{\rm vir}(t)}{M_{\rm vir}(t_{\rm gal})}\right )^{1/2},
\end{equation}
where $t_{\rm gal}=13.7$ Gyr is the age of the Galaxy, and  $v_c(R_\odot,t_{\rm gal})=220~{\rm km~s^{-1}}$ is the current circular velocity at the solar radius. $M_{\rm vir}(t)$ is adopted from the average fit reported in  \citet{griffen2016} from simulations of 24 Milky Way-sized halos. 
We adopt different values for $v_c(R_\odot)$ ranging from $5$--$220~{\rm km\,s^{-1}}$ that correspond to ages of $\sim 300$ Myr to the present age of $13.7$ Gyr.

For each $v_c(R_\odot)$, we simulate the motion of $(1-4)\times10^4$ NSBs according to their birth location, kick velocity, and DTD distribution described above. An NSB is considered to contribute to the Galactic enrichment if it merges within coordinates $R\leq R_{\rm max}$ and $|z| \leq z_{\rm max}$. In order to calculate the fraction $f_{\rm ret}(t)$ of NSBs retained by the Galaxy at different times, we use fixed values of $R_{\rm max}=20$ kpc and $z_{\rm max}=5$ kpc. The particular choice of $ R_{\rm max}$ roughly corresponds to the sum of the maximum value of $R_{\rm b}$ and the typical radial scale length of $\sim 3-4$ kpc. The value of $z_{\rm max}$ was taken to be the sum of the typical disk thickness $\sim 3$ kpc found in simulations by \citet{minchev2013} and  the typical remnant radius of a BNSM that explodes a few kiloparsecs above the disk plane. The latter is estimated to be $\sim 2$--$3$ kpc \citep{Thornton+1998} by using BNSM kinetic energy of $\sim 10^{51}$ erg and an ambient density of  $\sim 10^{-4}~{\rm cm}^{-3}$ that is expected at a height of a few kiloparsecs from the disk plane \citep{Miller2013}.
For $\eta(R,t)$ and $\beta^{\rm eff}(R,t)$, we first focus on the values $\eta_\odot(t)$ and $\beta^{\rm eff}_\odot (t)$ for the solar ring
defined by $7 \leq R \leq 9$ kpc. The values of $\eta_\odot$ are calculated using Eq.~\eqref{eq:eta} where only NSBs that merge within $z_{\rm max}$ are considered. For $\beta_{\odot}^{\rm eff}$, we first compute the average merger time of BNSMs within the solar ring that have $|z|<z_{\rm max}$. We then use the value of the average merger time to find the corresponding $\beta_{\odot}^{\rm eff}$ by assuming a DTD $\propto t^{\beta^{\rm eff}_\odot}$.

Figure~\ref{fig:minchev}(b) shows the evolution of $f_{\rm ret}$ for the entire Galaxy as well as $\eta_\odot$ and $\beta^{\rm eff}_\odot$ for the solar ring. At early times  ($t\lesssim 1$ Gyr),
$f_{\rm ret}$ ranges from 40\% to 50\%  but increases to $\sim 90\%$ by $t\sim 4$ Gyr. This is similar to the values obtained by \citet{safar2017} who considered a purely dark matter halo with NSBs traveling along the radial direction, and with a different criteria for deciding whether a BNSM contributes to the Galactic enrichment. The value of $\beta_{\odot}^{\rm eff}$ is always lower than the true  
value of $\beta=-1$, as expected. The $\beta_{\odot}^{\rm eff}$ is more negative at earlier times due to the higher escape rate of BNSMs, and approaches $-1$ at later times.

The evolution of $\eta_\odot$ on the other hand, is noticeably different from $f_{\rm ret}$ and $\beta_{\odot}^{\rm eff}$. It peaks at early times at $t\simeq t_0$ with a value reaching $\simeq 1.8$ 
and decreases with time for $t>t_0$. This is a direct consequence of the SFR from \citet{minchev2013} based on the inside-out formation of the Galaxy. As can be seen from Fig.~\ref{fig:minchev}(a), the surface SFR is higher at the center but drops sharply with $R$ for the first $\sim0.5$ Gyr after $t_0$, with a typical scale length of $\lesssim 1.5$ kpc. Subsequently, the scale length increases to $\gtrsim 3$ kpc within $\sim 2$ Gyr of disk evolution.
As a result of the steep drop of SFR with $R$ during the first $\sim 0.5$ Gyr of disk evolution, a substantial number of BNSMs that were originally born closer to the center of the Galaxy merge within the solar ring. 
This can be clearly seen in Fig.~\ref{fig:minchev}(c), which shows the distribution of the birth radii $R_{\rm b}$ of BNSMs that merge within the solar ring at two different stages of the disk evolution. 
At early stages of the disk evolution ($t_d\lesssim 0.5$ Gyr), $\gtrsim 50\%$ of the BNSMs that merge within the solar ring originated from $R\leq 4$ kpc, with the peak of the distribution of $R_{\rm b}$ at $R=1$--$3$ kpc. 
As the scale length increases with time, at $t_d\sim 4.5$ Gyr, only $\sim 5\%$ of BNSMs that originated from $R\leq 4$ kpc contribute to the solar ring, and the peak of the distribution lies within the ring at $R=7$--$9$ kpc. 
Because the radial dependence of the SFR stays constant for $t\leq t_0$, $\eta_\odot$ decreases slightly with decreasing values of $t$ as an increasing fraction of NSBs are able to escape due to the shallower Galactic potential.

An important point to note here is that for a given distribution of natal kick velocities, only $\eta_\odot$ is uniquely sensitive to the radial distribution of SFR, whereas $f_{\rm ret}$ and $\beta_{\odot}^{\rm eff}$ are mostly sensitive to the Galactic potential. 
To illustrate this, we calculate $f_{\rm ret}$, $\eta_\odot$, and $\beta_{\odot}^{\rm eff}$ assuming a surface SFR with a fixed radial scale length $R_{\rm d}=3$ kpc, i.e, $\Sigma \propto e^{-R/R_{\rm d}}$,
throughout the Galactic evolution and show the results in Fig.~\ref{fig:minchev}(d). 
When compared to Fig.~\ref{fig:minchev}(b), it can be seen clearly that whereas 
the evolution of $f_{\rm ret}$ and $\beta_{\odot}^{\rm eff}$ remain qualitatively unchanged, the evolution of $\eta_\odot$ changes dramatically. In this case, the evolution of $\eta_\odot$ becomes
very similar to $f_{\rm ret}$ which increases gradually instead of decreasing with time.

\section{Impact of Natal Kicks on GCE Calculations}
\begin{figure}[h]
\centerline{\includegraphics[width=84mm]{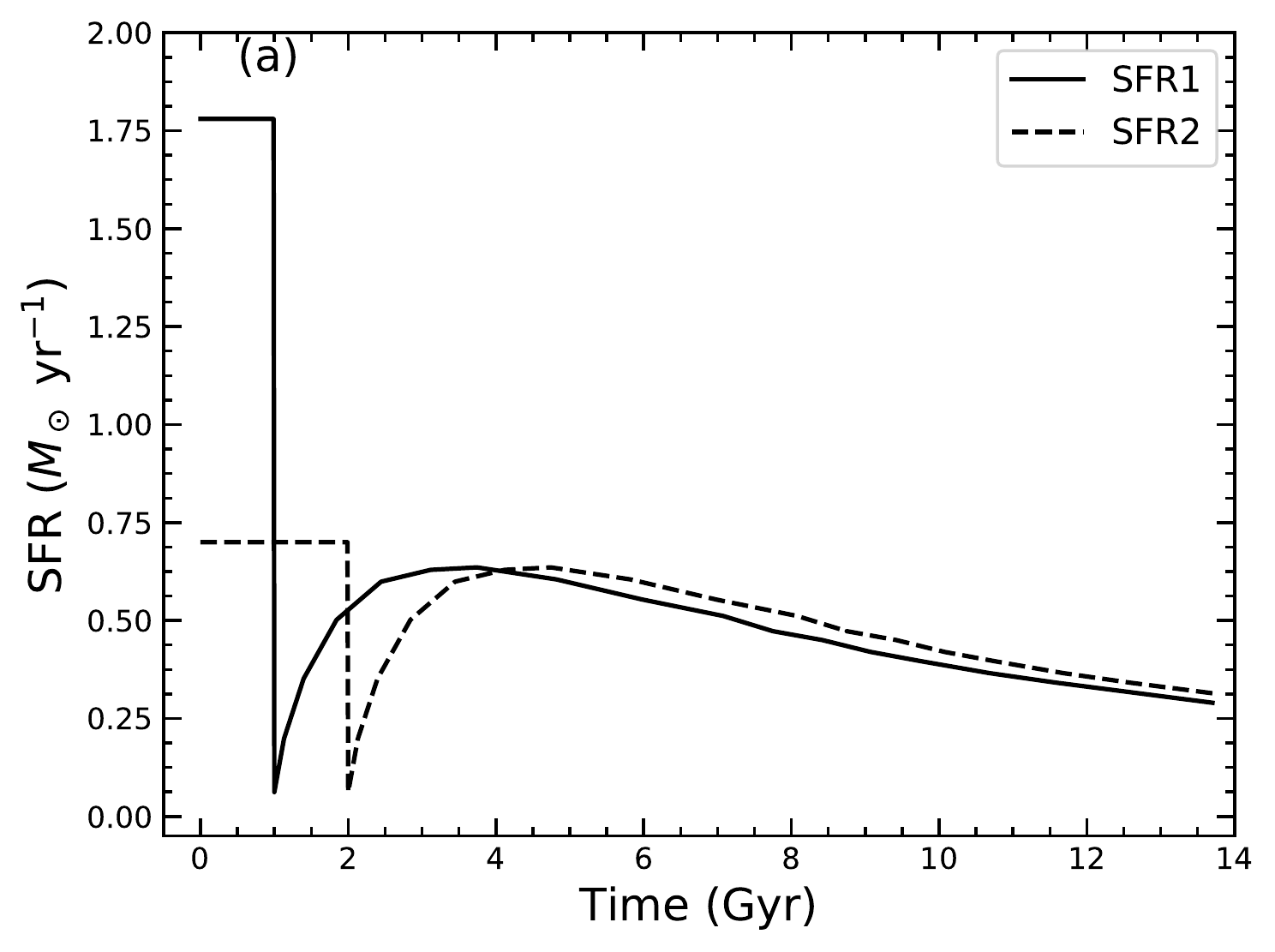}}
\centerline{\includegraphics[width=85mm]{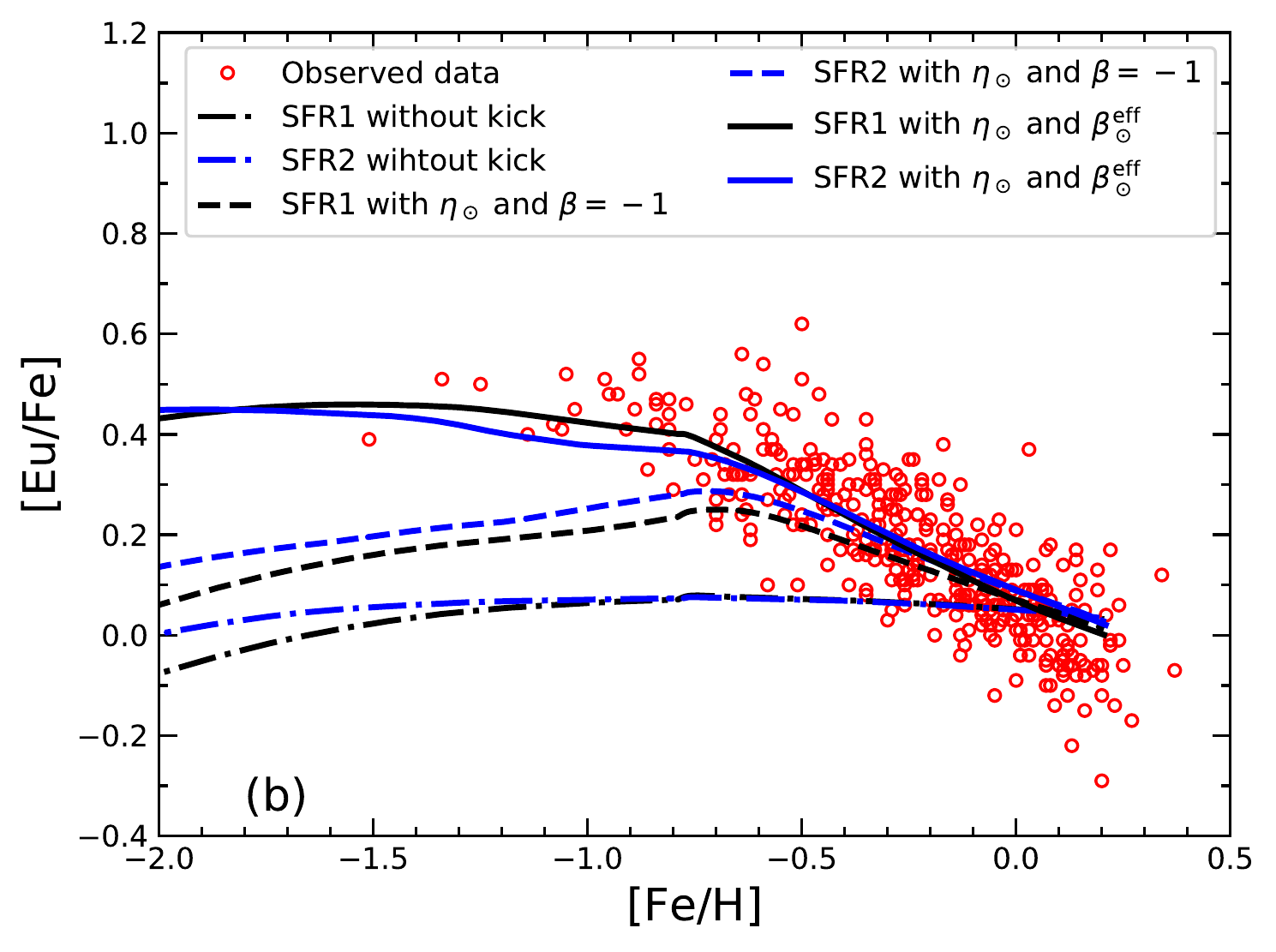}}
\centerline{\includegraphics[width=85mm]{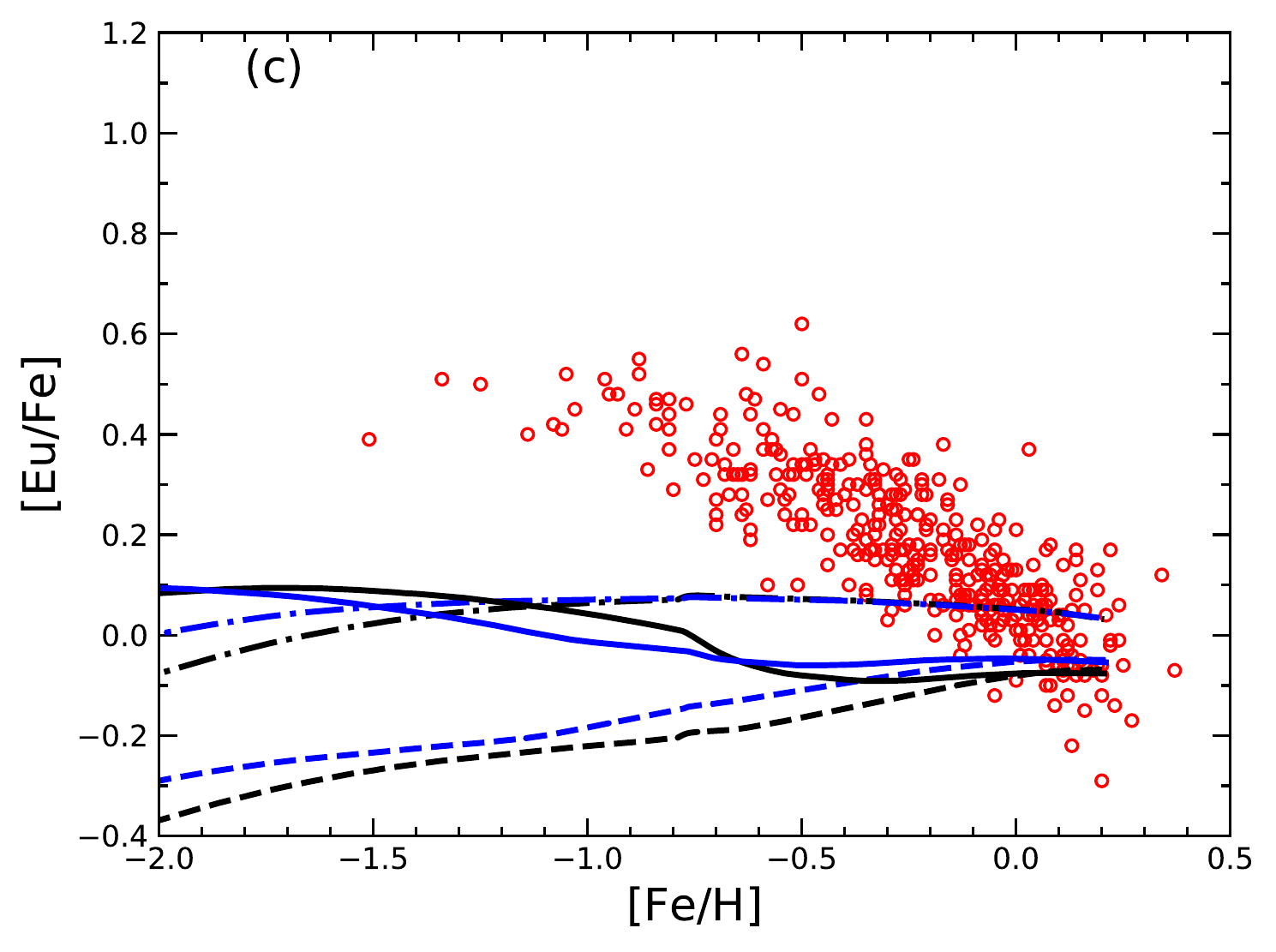}}
\caption{(a) Star formation rates SFR1 and SFR2. 
(b) Evolution of [Eu/Fe] with [Fe/H] for SFR1 and SFR2 
for cases including $\eta_\odot$ and $\beta^{\rm eff}_\odot$, including only $\eta_\odot$ ($\beta=-1$), and without including any kick effects.
(c) Same as (b), but with  $\eta_\odot$ and $\beta^{\rm eff}_\odot$ from Fig.~\ref{fig:minchev}(d), 
calculated using a fixed scale length of 3 kpc (see the text). In all cases $m^{\rm BNSM}_{\rm ej}=1.3\times10^{-2}\,M_\odot$ is used. Observational data are adapted from \citet{battistini2016}.}
\label{fig:minchev_gce} 
\end{figure}

In order to study the impact of natal kicks on GCE, we use the one-zone chemical evolution code \textsc{omega} \citep{omega2016,omega2017} to model the solar ring centered at $R_\odot=8$ kpc with a thickness of $2$ kpc, and take the closed box approximation for simplicity. The code tracks the contributions from low-  and intermediate-mass stars taken from \cite{karakas2010} as well as massive stars taken from \citet{kobayashi2006}, where $50\%$ of stars from $20$--$40\, M_\odot$ are assumed to explode as hypernovae.
Stars with initial masses between $3$--$8\,M_\odot$, that form white dwarfs are assumed to contribute toward SNe Ia with a DTD $\propto t^{-1}$ and a minimum delay time 
of $\sim 40$ Myr (corresponding to the lifetime of an $8\,M_\odot$ star). The number of SNe Ia per unit of stellar mass formed, $N_{\rm Ia}$, is fixed at $2\times 10^{-3}$ with the SNe Ia yields adopted from \citet{Iwamoto+1999} (for more details see \citealt{omega2016} and \citealt{ritter2018}).  
A fraction $f_{\rm BNSM}$ of massive stars are assumed to lead to BNSMs with a DTD $\propto t^{\beta}$ and ejecta mass $m^{\rm BNSM}_{\rm ej}$. 
We fix the value of $f_{\rm BNSM}=0.01$ and $\beta=-1$ when the effects due to natal kicks discussed in Sec.~\ref{sec:kick} are neglected.
The effects of natal kicks are included by replacing $f_{\rm BNSM}$ with $f_{\rm BNSM}\times \eta_\odot(t)$, and $\beta$ with $\beta^{\rm eff}_\odot(t)$, respectively.
The yields of \textsl{r}-process elements in the BNSM ejecta are 
assumed to follow the solar \textsl{r}-process pattern for mass number $A\geq 80$ from \citet{arnould2007}, which amounts to a mass fraction of $1.04\times 10^{-3}$ for Eu in the ejecta.
Because varying the ejecta mass $m^{\rm BNSM}_{\rm ej}$ amounts to an overall scaling of the Eu yield, its value is chosen between  $0.01$ and $0.02\, M_\odot$ that best fits to the data. 
We note that recent nucleosynthesis studies of BNSM outflows at different phases that produce a wide range of \textsl{r}-process nuclides \citep{Wanajo:2014wha,Just:2014fka,Wu:2016pnw,Siegel:2017nub} reported similar Eu yields as the number adopted here.

The SFR for $t\geq t_0$ in our calculation is adopted directly from \citet{minchev2013} using the surface SFR at 8 kpc that is multiplied by the area of the ring. It is known from existing calculations \citep[e.g.][]{Cote2019}, that with a constant value of $f_{\rm BNSM}$, i.e, neglecting the effect of natal kicks, the curve of [Eu/Fe] has a flat trend with [Fe/H]. Thus, when $f_{\rm BNSM}$ is replaced by $f_{\rm BNSM}\times \eta_\odot$, the curve of [Eu/Fe] is expected to follow the trend of $\eta_\odot$. Because the downward trend of [Eu/Fe] starts at [Fe/H]$\sim -0.8$, we assume that this coincides with $t=t_0$. For $t<t_0$, a constant value of SFR is chosen such that [Fe/H] reaches $\sim -0.8$ at $t=t_0$. Figure~\ref{fig:minchev_gce}(a) shows the resulting SFR for $t_0=1$ Gyr (SFR1) and $t_0=2$ Gyr (SFR2). The initial gas mass is calculated by requiring [Fe/H] to reach $0.2$ by the end of the evolution at $t=13.7$ Gyr, which gives values of  $10.5\times 10^9\, M_\odot$ and $9.4\times 10^9\, M_\odot$ for $t_0=1$ and $2$ Gyr, respectively. We note here that the resulting evolution of $\alpha$ elements, such as Mg, with the above choice of values agrees well with the observed trend.

In figure~\ref{fig:minchev_gce}(b), we show the evolution of [Eu/Fe] as a function of [Fe/H], using both SFRs, for three different cases that illustrate effects due to natal kicks: (i) completely neglecting the effect of natal kicks, i.e, including neither $\eta_\odot(t)$ nor $\beta^{\rm eff}_\odot(t)$ and using $\beta=-1$; (ii) including only $\eta_\odot(t)$ with $\beta=-1$; and (iii) including both $\eta_\odot(t)$ and $\beta^{\rm eff}_\odot(t)$.
When the effects of natal kicks are completely ignored, the flat trend of [Eu/Fe] for [Fe/H]$\gtrsim -1.5$ (see the dashed-dotted curves) consistent with the findings of \citet{Cote2019} is recovered.

In contrast, when only $\eta_\odot(t)$ is included with $\beta=-1$ in case (ii), it is clear that the trend of [Eu/Fe] follows that of $\eta_\odot(t)$ as expected. Specifically, [Eu/Fe] first increases with [Fe/H] for $t\leq t_0$ (corresponding to [Fe/H]$\lesssim -0.8$), and then decreases for $t>t_0$ ([Fe/H]$\gtrsim-0.8$). The slope for the decreasing curve for [Fe/H]$\gtrsim -0.8$, however, is slightly flatter than the observed data. Finally, when both $\eta_\odot(t)$ and $\beta^{\rm eff}_\odot(t)$ are included (solid curves) in case (iii), the smaller values of $\beta^{\rm eff}_\odot(t)<-1$, i.e, steeper DTD, help to counter the increasing values of $\eta_\odot$ for [Fe/H]$\lesssim-0.8$ yielding a flat curve. On the other hand, for $t>t_0$, as the values of $\beta^{\rm eff}_\odot$ continue to be lower than $-1$, it helps to steepen the slope of the [Eu/Fe] curve further that is primarily caused by the decreasing values of $\eta_\odot$. Overall, this leads to a very good agreement with
the observed trend. We note that $m^{\rm BNSM}_{\rm ej}=1.3\times10^{-2} M_\odot$ is chosen here to match the observed data for the case when both $\eta_\odot$ and $\beta^{\rm eff}_\odot$ are included.

From the above discussion, it is evident that both $\eta_\odot$ and $\beta_\odot^{\rm eff}$ are important for the evolution of [Eu/Fe]. In particular, the decreasing trend of $\eta_\odot$ for $t>t_0$ due to the inside-out formation of the MW disk is crucial in producing the decreasing trend in [Eu/Fe] for [Fe/H]$\gtrsim -0.8$. 
To reinforce this, we perform additional GCE calculations using the values of $\eta_\odot$ and $\beta_\odot^{\rm eff}$ with a fixed scale length of 3 kpc shown in Fig.~\ref{fig:minchev}(d), and show the resulting [Eu/Fe] evolution in Fig.~\ref{fig:minchev_gce}(c).
In this case, the monotonically increasing $\eta_\odot$ results in an increasing trend of [Eu/Fe] when $\beta=-1$ is used. When $\beta_\odot^{\rm eff}$ is included, it can at best counter the negative impact of $\eta_\odot$ to yield a flat curve for [Fe/H]$\gtrsim -0.8$ similar to calculations that neglect the effect of kick altogether (see Fig.~\ref{fig:minchev_gce}(c)). 

\section{Discussion and Conclusions}
\begin{figure}[ht]
\centerline{\hspace*{-.2cm}\includegraphics[width=82mm]{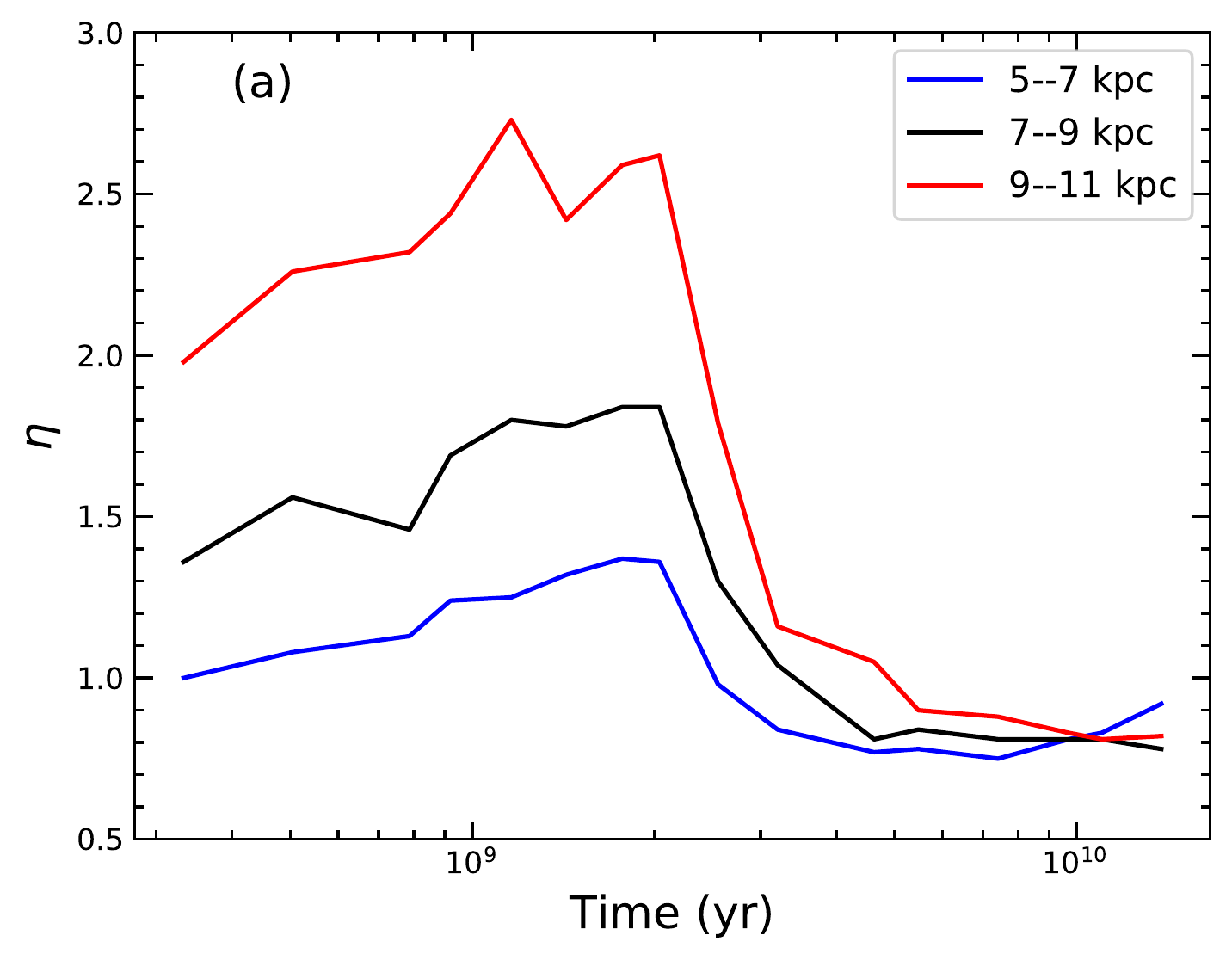}}
\centerline{\hspace*{-.4cm}\includegraphics[width=84mm]{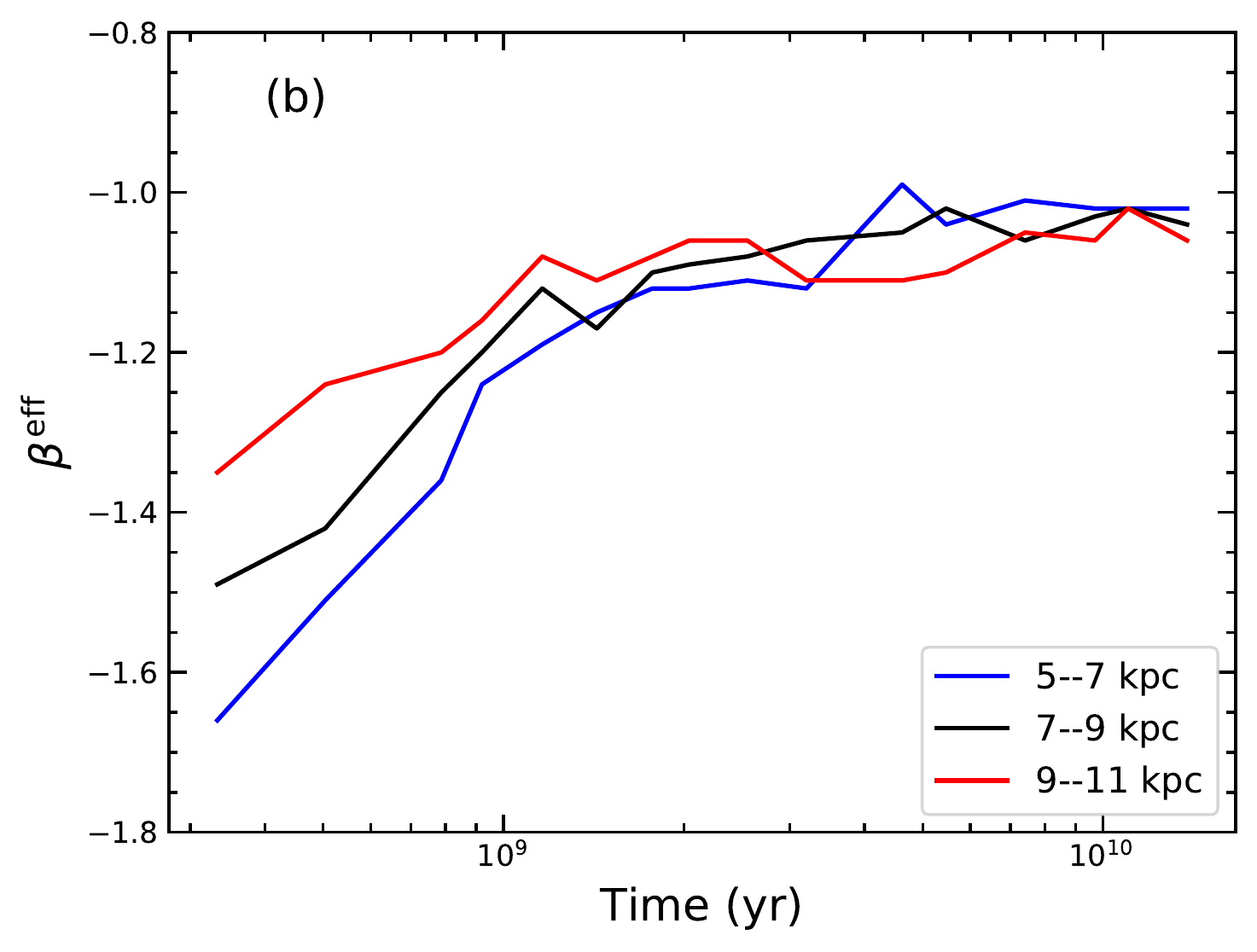}}
\caption{(a) Evolution of $\eta$ with $t_0=2$ Gyr for rings at different distances from the center. (b) Corresponding evolution of $\beta^{\rm eff}$.}
\label{fig:eta}
\end{figure}

In this Letter we studied the effects of natal kicks of NSBs on the GCE of elements like Eu that are almost entirely produced by the \textsl{r}-process, assuming that BNSMs are the sole source of \textsl{r}-process with a standard DTD $\propto t^{-1}$.
We find that natal kicks alter both the effective DTD and the occurring frequency of BNSMs in the solar neighborhood. 
In particular, the effect is amplified when the birth locations of NSBs are sampled according to an SFR that is consistent with the inside-out formation of the MW disk. 

During the first $\sim 1$ Gyr of the disk formation, the solar vicinity gets a large enhancement in the effective BNSM frequency from NSBs that originate from the inner parts of the disk but merge within the solar neighborhood 
due to kick-induced migration. This enhancement decreases with time and gives rise to the decreasing [Eu/Fe] trend for [Fe/H] $\gtrsim -0.8$. Additionally, natal kicks also lead to an effective DTD steeper than $\propto t^{-1}$, which further helps steepen the curve for [Eu/Fe] versus [Fe/H] for [Fe/H] $\gtrsim -0.8$ as well as flatten it for [Fe/H]$\lesssim -1$.
When these two effects of natal kicks, namely, $\eta_\odot$ and $\beta^{\rm eff}_\odot$, are combined together with  the inside-out formation of the MW disk, the decreasing [Eu/Fe] at [Fe/H]$\gtrsim -0.8$ can be naturally reproduced  without the need of additional \textsl{r}-process sources or modifications to the DTD $\propto t^{-1}$ for BNSM.

Because the exact evolution of [Eu/Fe] could depend on the assumed kick velocity distribution or the minimum merger time $t_{\rm merge}^{\rm min}$, we also performed additional calculations taking different values of $v_0$ and the minimal kick velocity, as well as the $t_{\rm merge}^{\rm min}$ to test the robustness of our conclusion.
The corresponding results in Appendices~\ref{app:kick} and \ref{app:mergertime} clearly show that the decreasing trend of [Eu/Fe] is not very sensitive to these parameters and the overall trend that matches the data well can be similarly reproduced. The effect of changing the minimum delay time for SN 1a is explored in Appendix~\ref{app:SN1a} which shows that the results are essentially unchanged. 

Although we adopted a simple one-zone closed box model for the GCE using \textsc{omega} for the solar vicinity, similar calculations using the same code have shown the mean trends are captured well when compared to more sophisticated GCE calculations \citep{Cote2019}. We also explored the effects of including outflow and inflow and found that they have a negligible impact on the results (see Appendix~\ref{app:outflow}).
Nevertheless, detailed calculations for the chemodynamical evolution of the MW that take into account the migration of NSBs due to natal kicks are highly desirable. Such calculations, however, are likely to be computationally demanding and expensive.

An interesting consequence of natal kicks is that the evolution of $\eta(R,t)$ in particular has a strong radial dependence and is thus different for each ring. Figure~\ref{fig:eta} shows the evolution of $\eta$ and $\beta^{\rm eff}$ for $t_0=2$ Gyr for different rings where the strong radial dependence of $\eta$ is evident in contrast to $\beta^{\rm eff}$, which is roughly similar for all rings. 
As can be seen from the figure, the maximum value of $\eta_{\rm max}$ that occurs at $t\approx t_0$ as well as the value of $\eta$ at any given time is lower for rings closer to the Galactic center. Because $\eta$ directly impacts the value of [Eu/Fe], its value is also expected to be lower for rings closer to the center and vice versa. Additionally, for $t>t_0$, the slope of $\eta$ becomes flatter (steeper) for rings closer (farther) to the Galactic center. Although the exact trend would depend on the details of the SFR and gas mass relevant for the ring, this is expected to result in a flatter (steeper) downward slope of [Eu/Fe] versus [Fe/H] for rings closer (farther) than the solar neighborhood. Such a prediction, in principle, can be verified if Eu is measured in a sufficient number of stars over a range of [Fe/H] at other locations of the MW.

Other effects that have already been shown to be important in explaining the metallicity distribution are radial gas flows and migration of stars in the disk \citep[see e.g.][]{Schonrich09a,Schonrich09b,minchev2013, Schonrich17}.  Although the details are complicated and beyond the scope of this paper, the net result of such mixing is that it tends to weaken the radial metallicity gradient slightly. Clearly, this will also impact the [Eu/Fe] trend for the solar neighborhood due some amount of mixing of stars and gas from the inner (outer) regions with slightly flatter (steeper) [Eu/Fe]  versus [Fe/H] curve  from inner (outer) region. On the other hand, because of the radial metallicity gradient, at any given time, the value of [Fe/H] will be higher in the inner region and lower in the outer region. Thus, for the solar ring, radial mixing will bring in stars with lower [Eu/Fe] from the inner regions but with higher [Fe/H]. Exactly the opposite is applicable for stars coming from the outer region. Whereas the impact of stellar migration can only be calculated with detailed chemodynamical calculations, effectively, this would lead to some scatter around the mean trend \citep{Tsujomoto2019}.

Finally, our results show that BNSMs with a DTD $\propto t^{-1}$ alone is sufficient to explain both the origin as well as the evolution of \textsl{r}-process elements in the Galaxy for [Fe/H]$\gtrsim -2$. This, however, does not rule out other sources for \textsl{r}-process, but rather indicates that their contribution is likely subdominant compared to BNSM. We note, however, that this conclusion does not apply to the origin of \textsl{r}-process observed in very metal-poor stars formed in the early Galaxy where additional sources could still be required \citep{Wehmeyer2015}.

\acknowledgements
The authors thank Benoit C\`ote for help with using \textsc{omega}, Ivan Minchev for his valuable inputs and comments, and the anonymous referee for a detailed and constructive report.
M.-R.W. acknowledges support from the Academia Sinica by grant No.~AS-CDA-109-M11, Ministry of Science and Technology, Taiwan under grant No.~108-2112-M-001-010, and the Physics Division, National Center of Theoretical Science of Taiwan.
Z.Y. acknowledges the support from Special Funding for Advanced Users through by the LAMOST FELLOWSHIP and the Shanghai Sailing Program (Y955051001).

\appendix

\section{Dependence on Natal Kick Velocity Distribution}\label{app:kick}

\begin{figure*}[h]
\centerline{\hspace*{.30cm}\includegraphics[width=78mm]{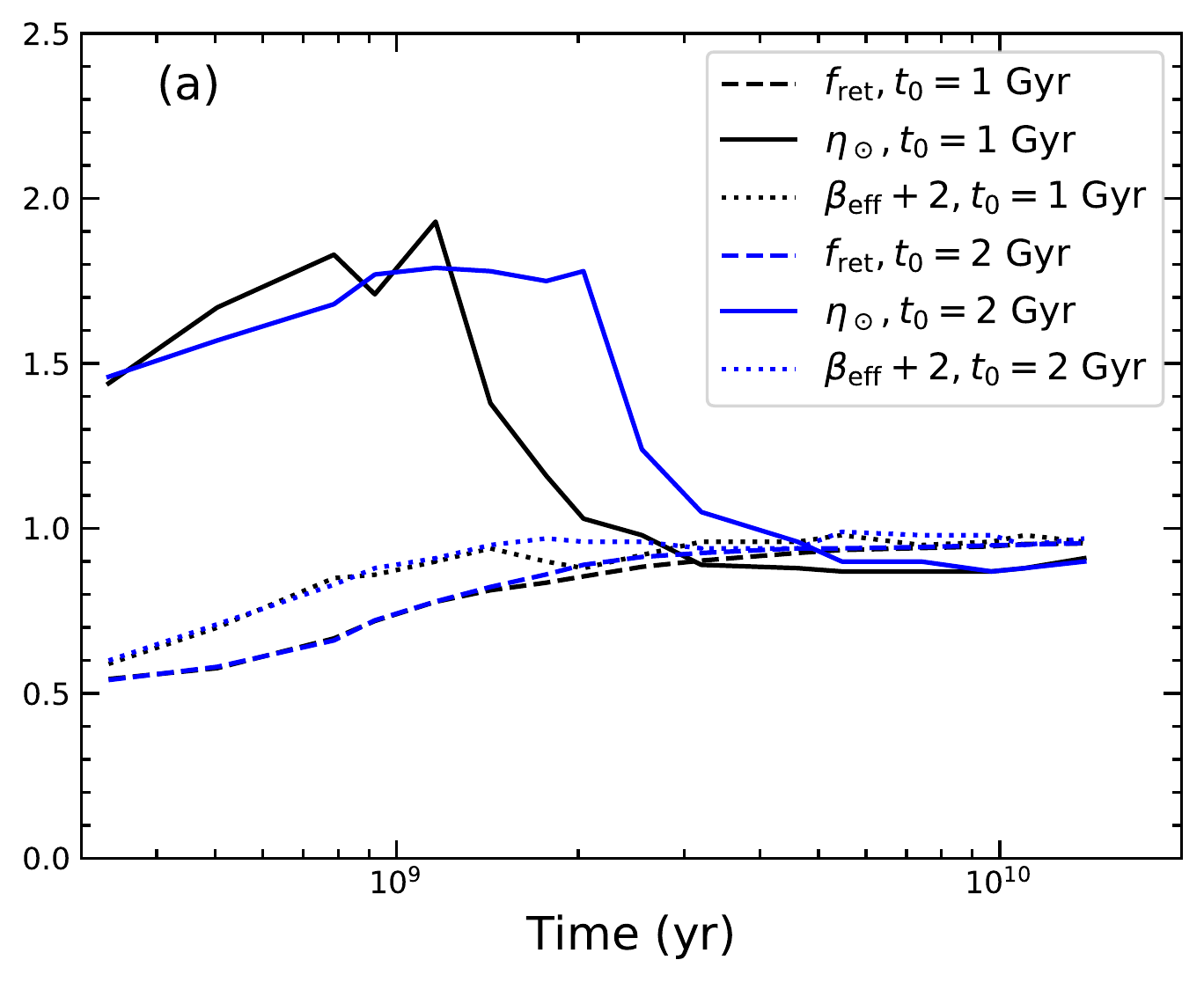} \includegraphics[width=85mm]{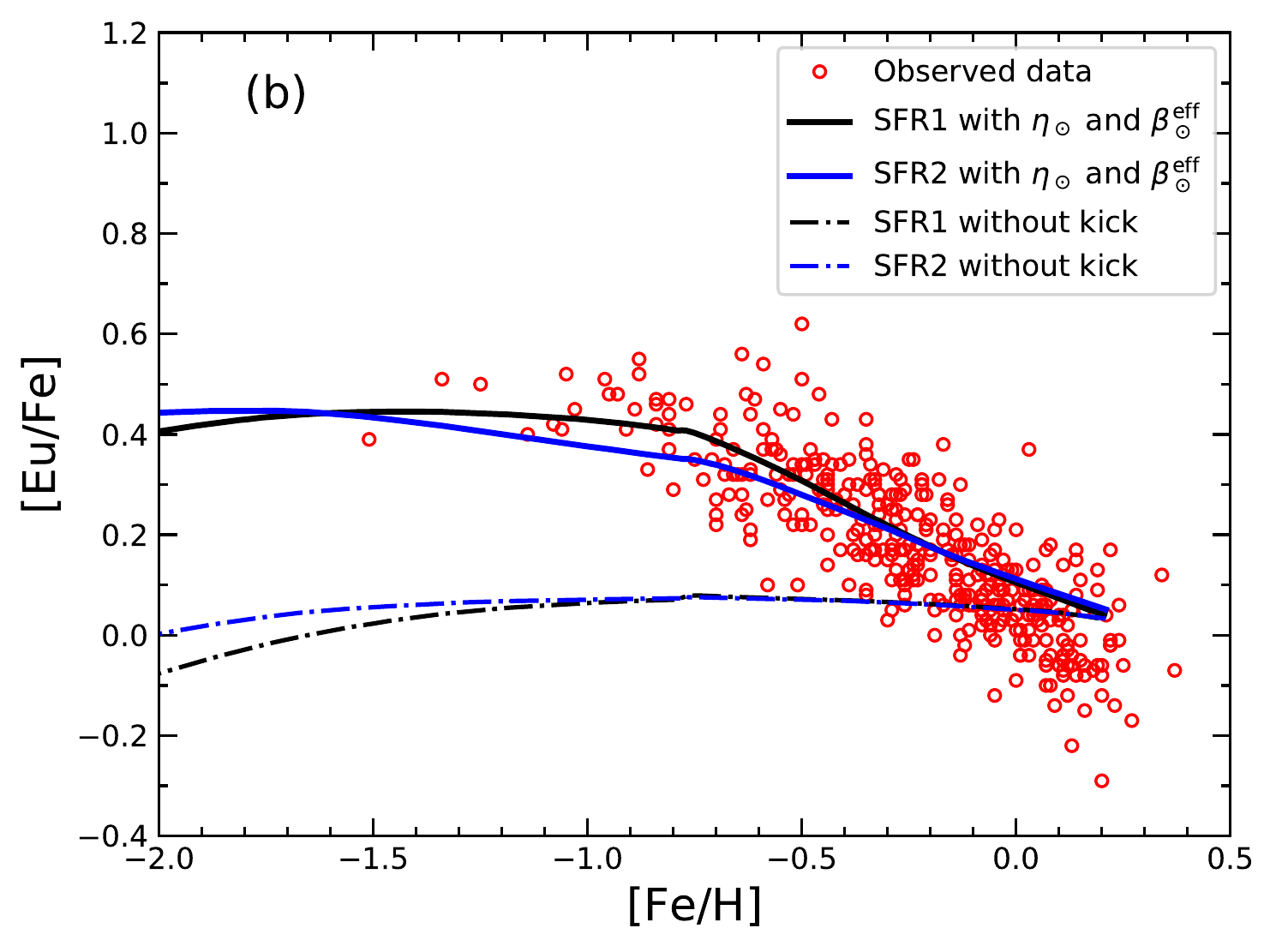}}
\centerline{\hspace*{.30cm}\includegraphics[width=78mm]{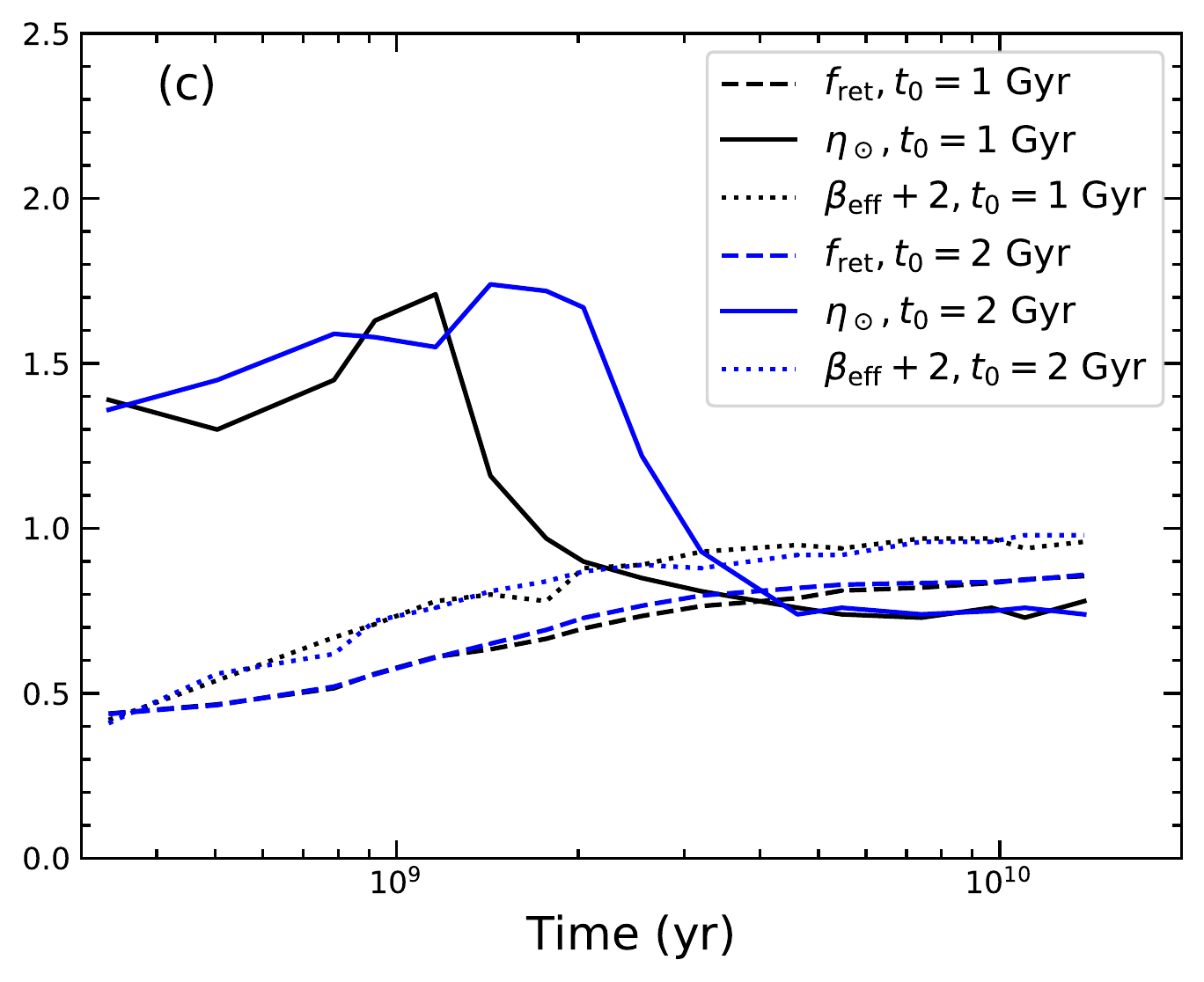} \includegraphics[width=85mm]{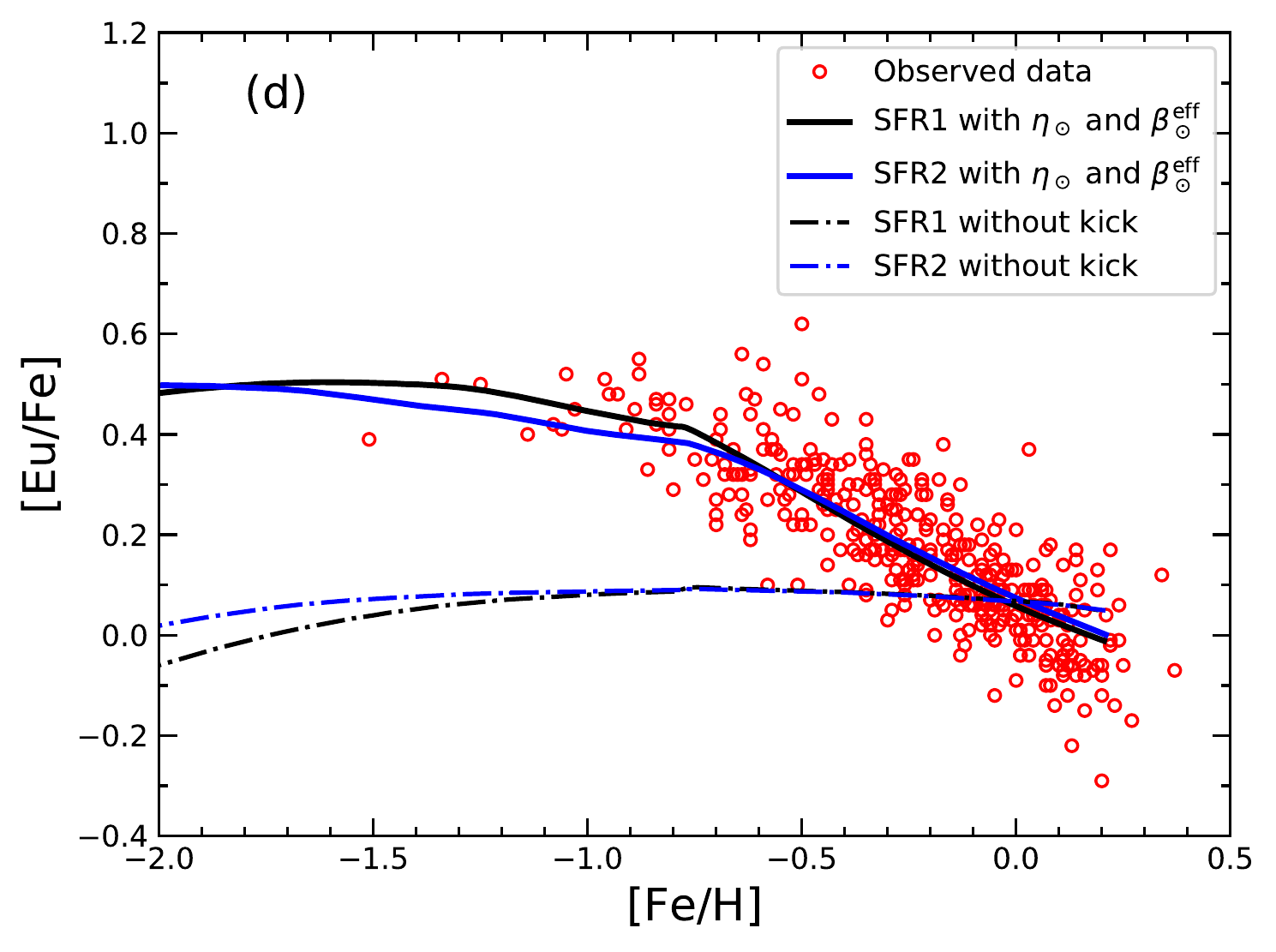}}
\caption{Effect of changing the distribution of $v_{\rm kick}$. (a) Same as Fig.~\ref{fig:minchev}(b), but with $v_0=60~{\rm km~s^{-1}}$, (b) Same as Fig.~\ref{fig:minchev_gce}(c), but with $v_0=60~{\rm km~s^{-1}}$, (c) Same as (a), but with $v_0=120~{\rm km~s^{-1}}$, (d) Same as (b), but with $v_0=120~{\rm km~s^{-1}}$ with $m^{\rm BNSM}_{\rm ej}=1.35\times10^{-2}\,M_\odot$.}
\label{fig:vkick}
\end{figure*}

The results presented in the main text are computed using the same distribution for the kick velocity $\propto \exp(-v/v_0)$ with $v_0=90~{\rm km~s^{-1}}$ with a minimum value of $10~{\rm km~s^{-1}}$. 
Here, we explore the effect of changing the distribution of kick velocity on the results. First, we repeat our calculations using $v_0=60~{\rm km~s^{-1}}$ and $120~{\rm km~s^{-1}}$. 
The results are qualitatively similar, as can be seen from Fig.~\ref{fig:vkick}.
The slope of [Eu/Fe] for [Fe/H]$\gtrsim -0.8$
is slightly flatter (steeper) for lower (higher) average kick velocity. 
The slight change in the slope is caused by slightly higher values of $\beta_\odot^{\rm eff}$ for lower average kick velocity and vice versa. On the other hand, for lower average kick velocity, 
the NSBs migrate slightly less, but fewer of them escape from the potential. 
The two effects counterbalance each other such that $\eta_\odot$ remains roughly unchanged. 
Overall, the results are only weakly sensitive to the average kick velocity, with higher values resulting in even better fits to the observed data.

\begin{figure*}[h]
\centerline{\hspace*{.30cm}\includegraphics[width=78mm]{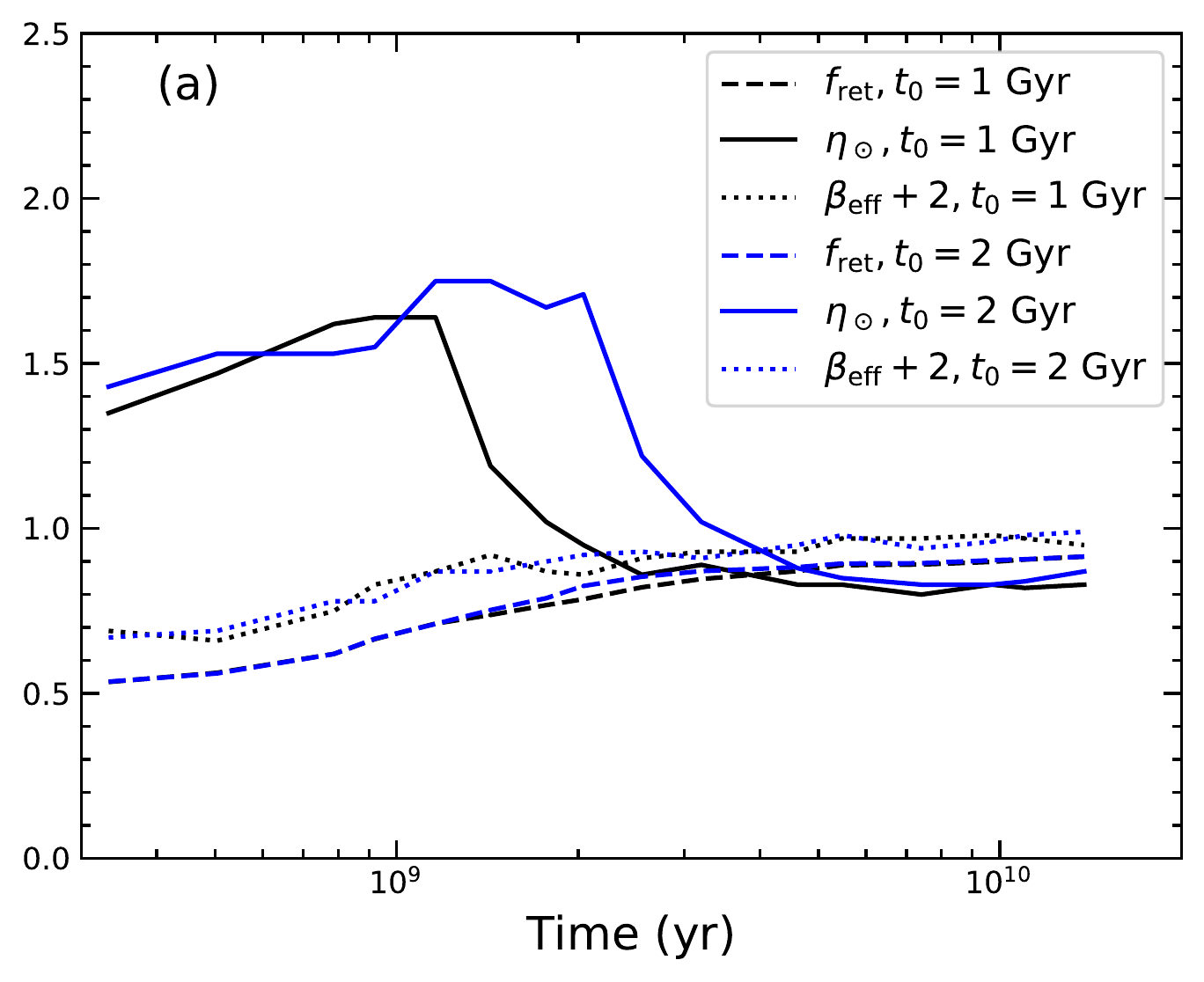} \includegraphics[width=85mm]{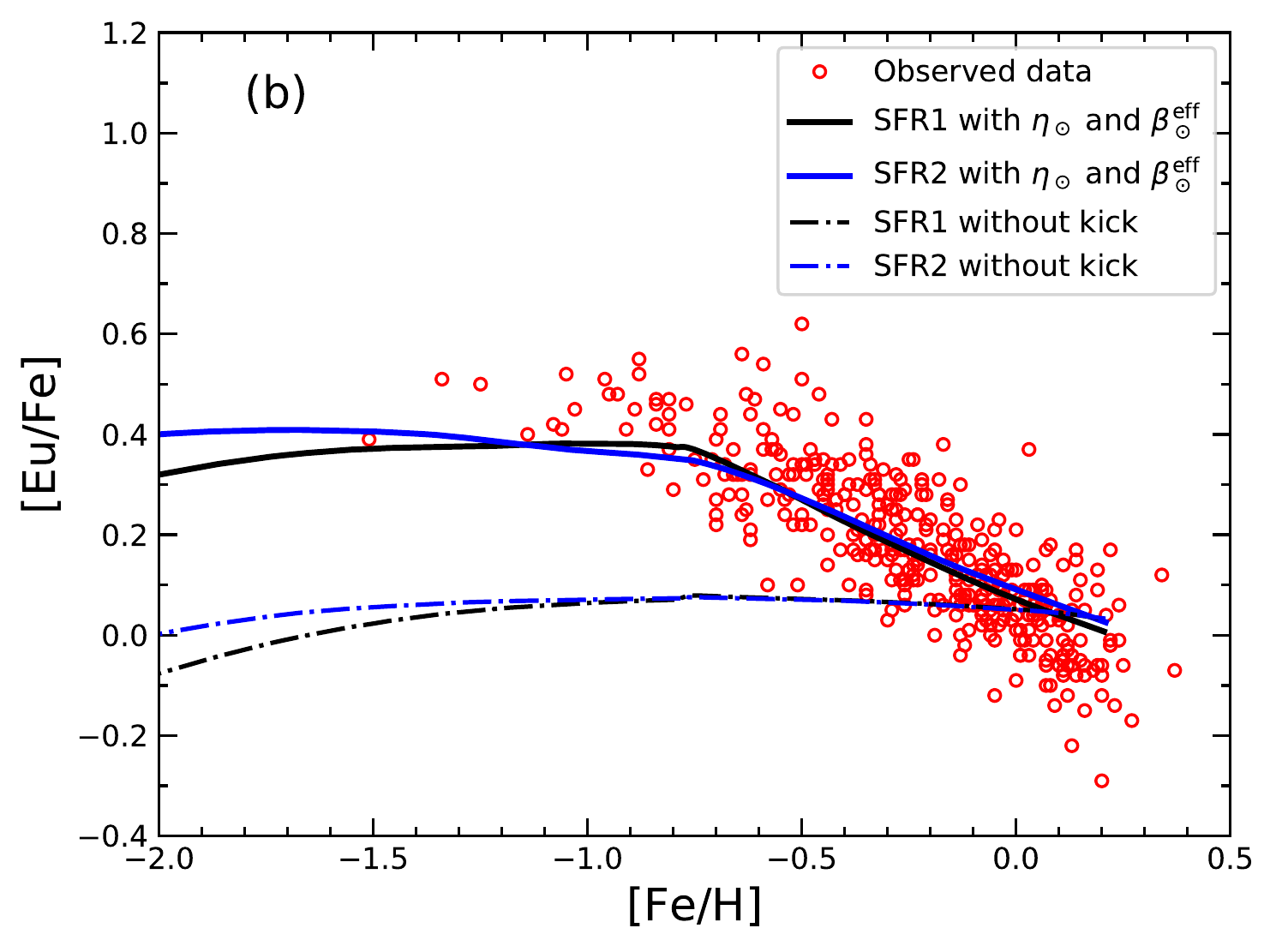}}
\caption{(a) Same as Fig.~\ref{fig:minchev}(b), but with $v_{\rm min}=0~{\rm km~s^{-1}}$ . (b) Same as Fig.~\ref{fig:minchev_gce}(c), but with $v_{\rm min}=0~{\rm km~s^{-1}}$.}
\label{fig:vkickmin}
\end{figure*}

We further explore 
the effect due to the contribution of NSBs with lower kick velocities below $v\lesssim 10$~km~s$^{-1}$, that could arise from binaries that involve low-mass core-collapse SNe. We, again, repeat our calculations using $v_{\rm min}=0~{\rm km~s^{-1}}$  with $v_0=90~{\rm km~s^{-1}}$. 
Compared to the default case with $v_{\rm min}=10$~km~s$^{-1}$, the fraction of NSBs with $v\lesssim 20$~km~s$^{-1}$ is increased from $\sim 10\%$ to $\sim 20\%$.
As shown in Fig.~\ref{fig:vkickmin}, this only leads to a marginally lower values of $\eta_\odot$ and $\beta_\odot^{\rm eff}$ relative to the default case. 
Overall, the results qualitatively remain unchanged.

\section{Dependence on Minimum BNSM Merger Time}\label{app:mergertime}

\begin{figure*}[h]
\centerline{\hspace*{.30cm}\includegraphics[width=78mm]{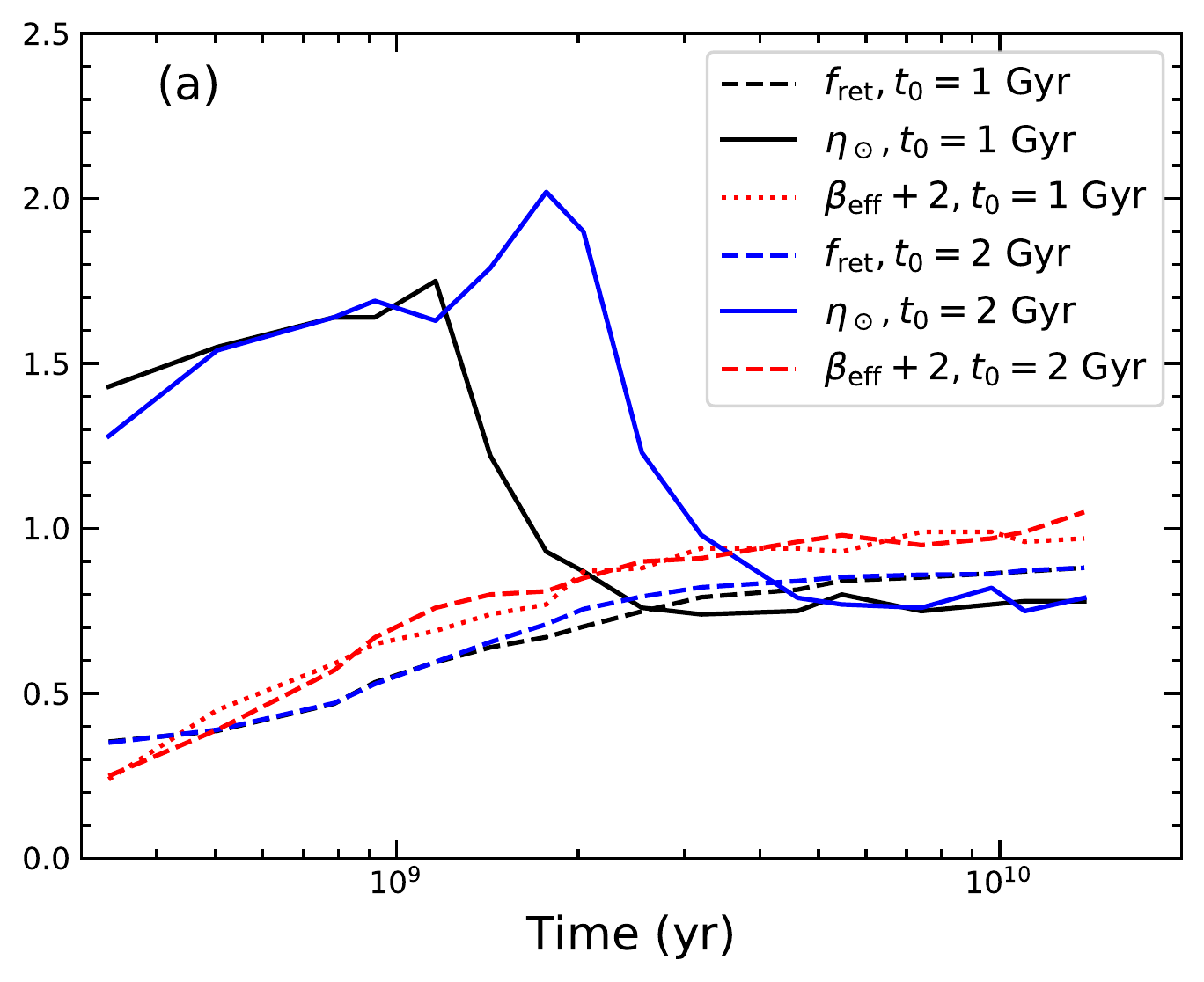}
\includegraphics[width=85mm]{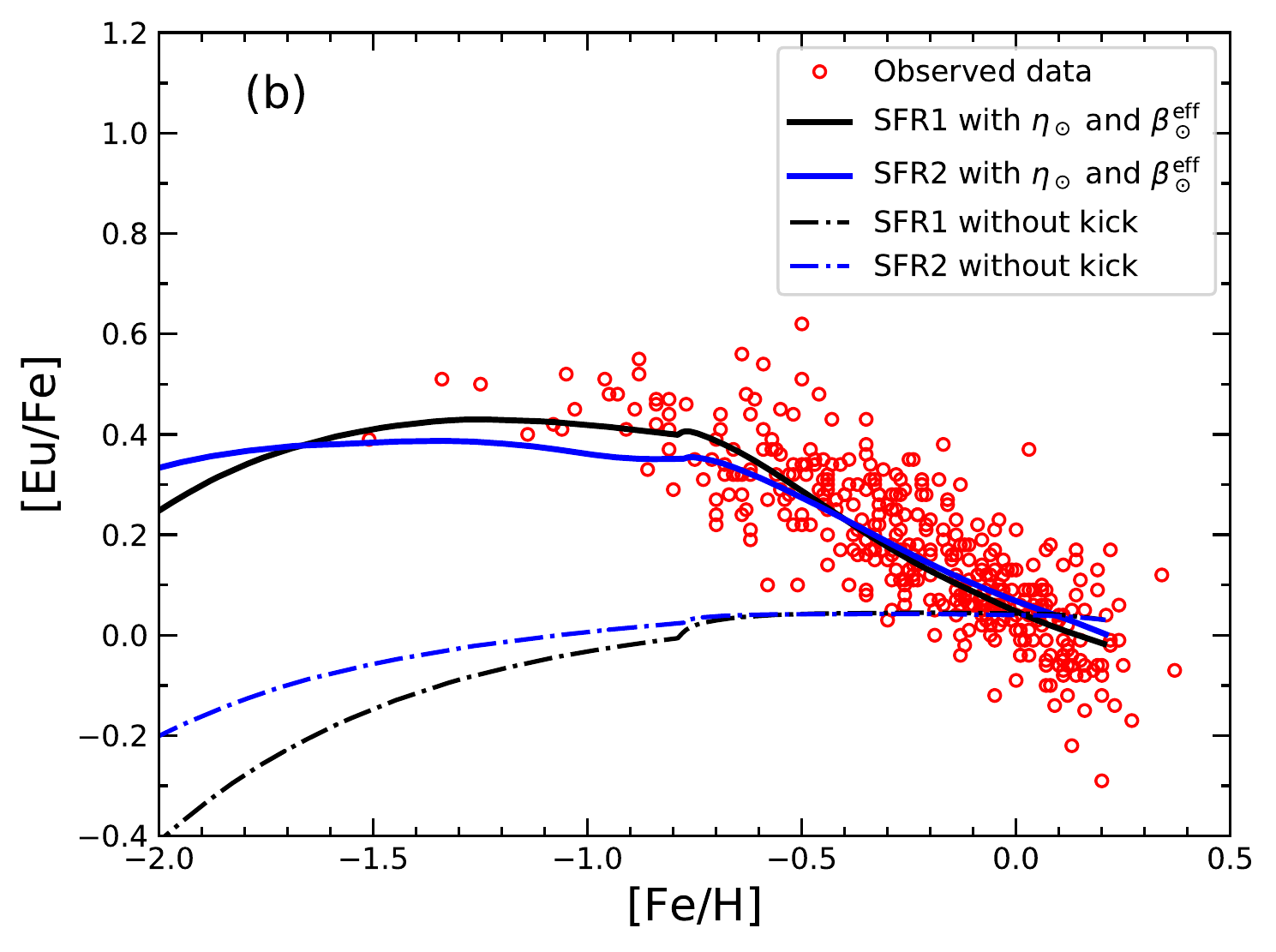}}

\caption{(a) Same as Fig.~\ref{fig:minchev}b but with $t_{\rm merge}^{min}=30$ Myr . (b) Same as Fig.~\ref{fig:minchev_gce}c but with $t_{\rm merge}^{\rm min}=30$ Myr.}
\label{fig:minchev_30Myr} 
\end{figure*}
We also explore the dependence
of our results on the choice of  $t_{\rm min}^{\rm merge}$, by repeating our calculations with $t_{\rm min}^{\rm merge}=30$ Myr. 
Figure~\ref{fig:minchev_30Myr} shows that the corresponding results are very similar to the calculations with $t_{\rm min}^{\rm merge}=10$ Myr shown in Fig.~\ref{fig:minchev_gce}. 
As before, $\beta_\odot^{\rm eff}$ helps to flatten the curve for [Fe/H]$\lesssim -0.8$ by countering the increasing values of $\eta_\odot$. 
Above [Fe/H]$\sim -0.8$, $\beta_\odot^{\rm eff}$  acts in tandem with the $\eta_\odot$ to produce a negative slope for [Eu/Fe] that matches the observed data very well. 
Thus, the results are not sensitive to the choice of $t_{\rm min}^{\rm merge}$.

\section{Dependence on Minimum delay time of SN Ia \label{app:SN1a}}
\begin{figure*}[h]
\centerline{\hspace*{.30cm} \includegraphics[width=85mm]{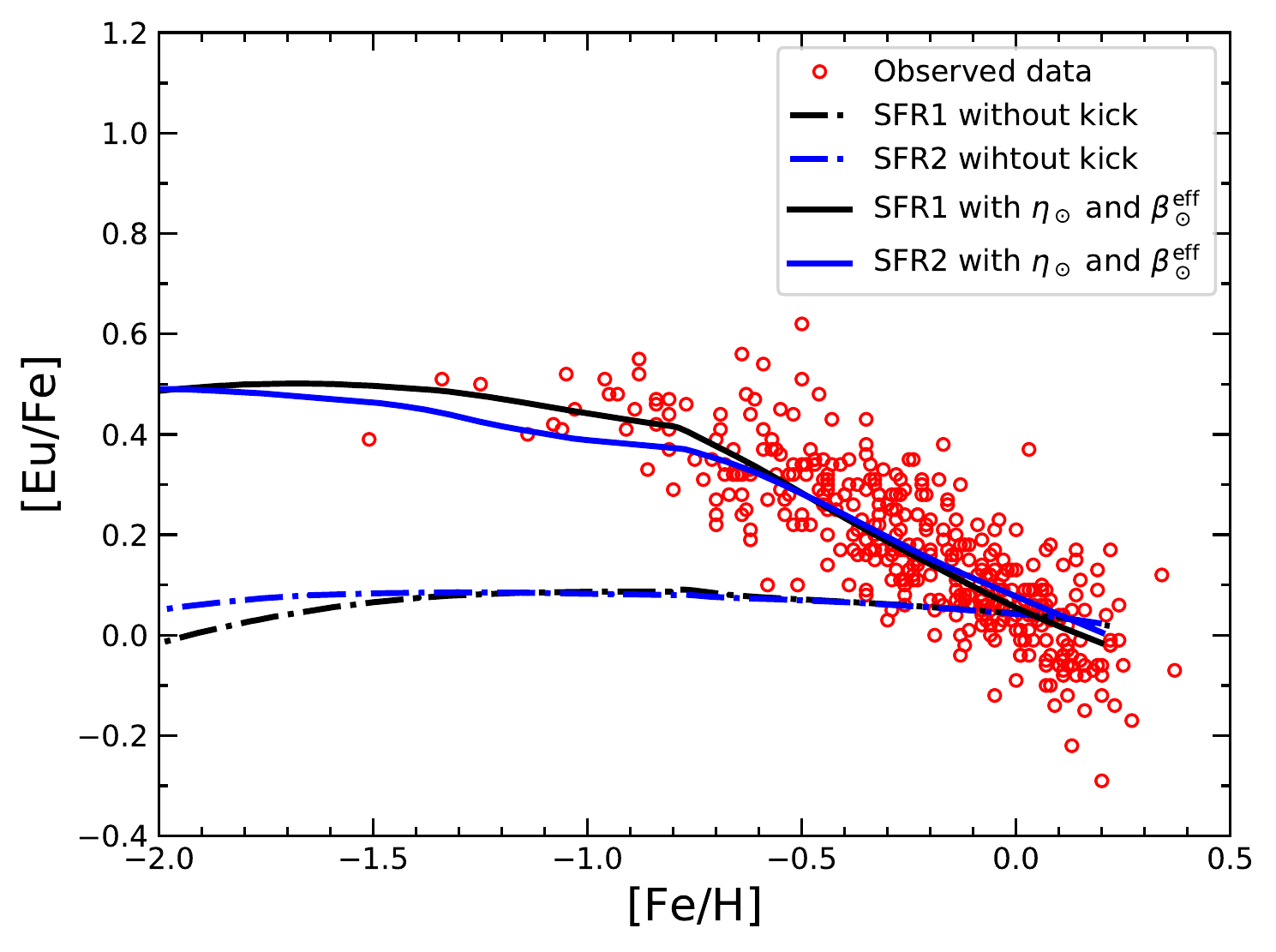}}
\caption{ Same as Fig.~\ref{fig:minchev_gce}(c), but with minimum SN Ia delay time of 400 Myr with $N_{\rm Ia}=10^{-3}$.} 
\label{fig:Ia}
\end{figure*}
As shown in \citet{hotokezaka+2018}, the 
minimum delay time of SN Ia can also affect the evolution of [Eu/Fe] versus [Fe/H]. 
In this appendix, we explore this effect by changing the minimum delay time from the default 40 Myr to 400 Myr, similar to the values considered in \citet{hotokezaka+2018}.
Figure~\ref{fig:Ia} shows the results where the parameter $N_{\rm Ia}$ is reduced to $10^{-3}$ in order to ensure the final value of [Fe/H]$\sim 0.2$ remains unchanged. The resulting curve for [Eu/Fe] versus [Fe/H] is almost unchanged compared to our default model that clearly shows that the results are not sensitive to the variation of the minimum delay time of SNe Ia.

\section{Dependence on Gas Inflow and Outflow}\label{app:outflow}
\begin{figure*}[h]
\centerline{\hspace*{.30cm} \includegraphics[width=85mm]{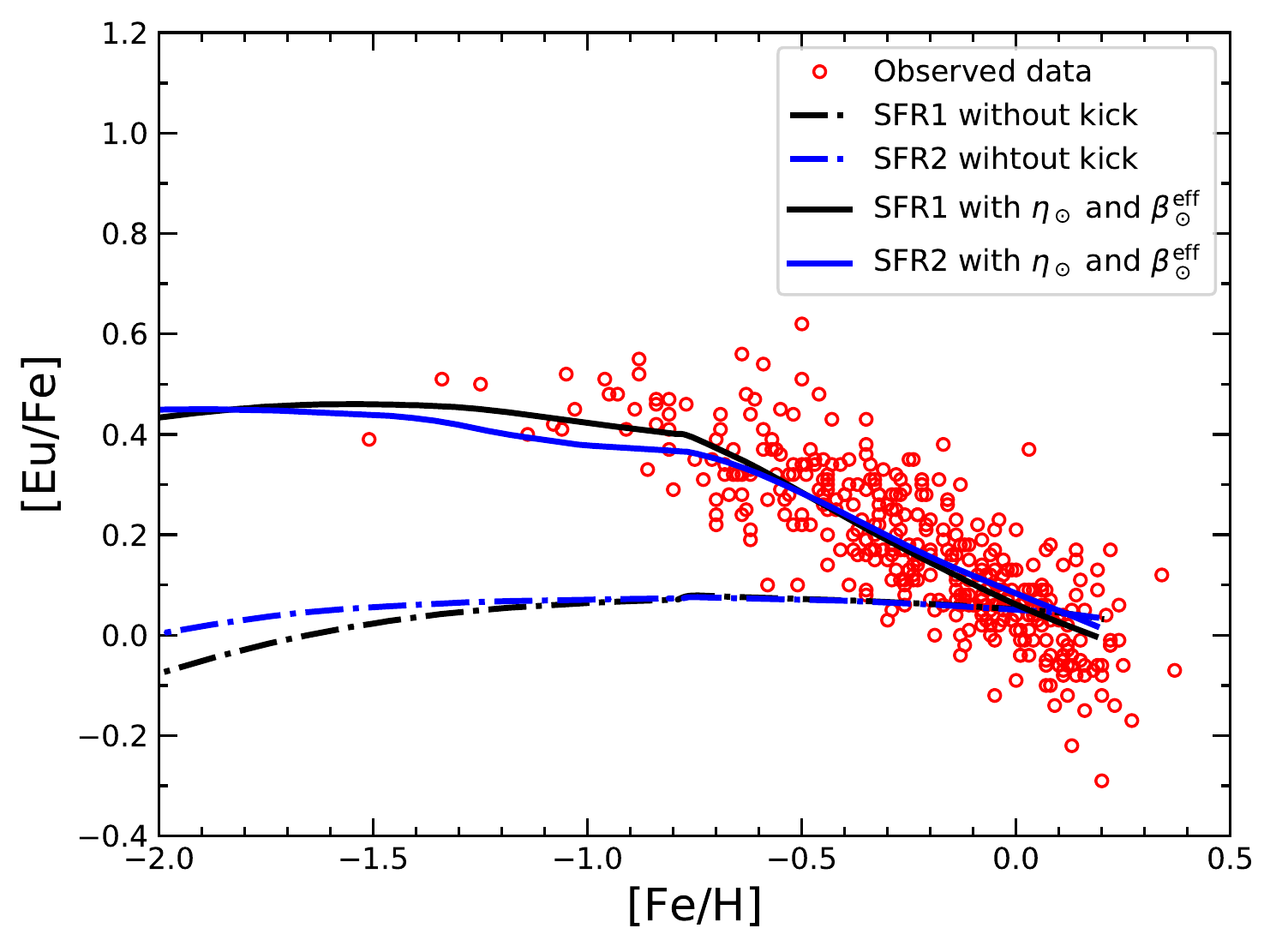}}
\caption{Same as Fig.~\ref{fig:minchev_gce}(c), but with inflow and outflow as described in Appendix~\ref{app:outflow}.}
\label{fig:outflow}
\end{figure*}
As mentioned in the main text, our results are based on a closed box model for the GCE. In this appendix, we explore the effect of including the outflow and inflow of gas on the [Eu/Fe] versus [Fe/H] trend. 
The outflow is assumed to be proportional to the SFR, whereas the inflow rate is assumed to be proportional to the outflow rate (see \citealt{omega2017} for details). 
The proportionality constant for the outflow rate and the ratio between the inflow and outflow rates are taken to 0.1 and 1, respectively. 
These values are consistent with numerical simulation of gas outflow and inflow of the Galactic disk \citep{kim2018}. 
We use the same initial gas masses of $10.5\, M_\odot$ and $9.4\, M_\odot$ for SFR1 and SFR2, respectively, as in the main text. 
The resulting trend of [Eu/Fe] versus [Fe/H] in this case, shown in Fig.~\ref{fig:outflow}, is similar to the closed box calculations, indicating that inflow and outflow do not affect the results.

\clearpage


\end{document}